\documentclass[twocolumn]{aastex631}
\usepackage{amsmath}

\begin{document}

\title{Extending the [C/N]-Age Calibration: Using Globular Clusters to Explore Older and Metal-Poor Populations}

\correspondingauthor{Peter Frinchaboy}
\email{p.frinchaboy@tcu.edu}

\author[0000-0003-4019-5167]{Taylor Spoo}
\affiliation{Department of Physics and Astronomy, Texas Christian University, TCU Box 298840 \\
Fort Worth, TX 76129, USA }

\author[0000-0002-2162-9911]{Katelyn Thomas}
\affiliation{Department of Physics and Astronomy, Emory University, 400 Dowman Dr, Atlanta, GA 30322}
\affiliation{Department of Physics and Astronomy, Texas Christian University, TCU Box 298840 \\
Fort Worth, TX 76129, USA }

\author[0009-0004-2667-9995]{Ellie ``Kaleo" Toguchi-Tani}
\affiliation{Department of Astronomy, Whitman College, 280 Boyer Ave, Walla Walla, WA, 99362 }
\affiliation{Department of Physics and Astronomy, Texas Christian University, TCU Box 298840 \\
Fort Worth, TX 76129, USA }

\author[0000-0003-2602-4302]{Jonah Otto}
\affiliation{Department of Physics and Astronomy, Texas Christian University, TCU Box 298840 \\
Fort Worth, TX 76129, USA }

\author[0000-0001-9738-4829]{Natalie Myers}
\affiliation{Department of Physics and Astronomy, Texas Christian University, TCU Box 298840 \\
Fort Worth, TX 76129, USA }

\author[0000-0002-4818-7885]{Jamie Tayar}
\affiliation{Department of Astronomy, University of Florida, 211 Bryant Space Science Center, Gainesville, FL 32611, USA}

\author[0000-0002-1043-8853]{Jessica S. Schonhut-Stasik}
\affiliation{Vanderbilt University, Physics \& Astronomy Department, 6301 Stevenson Center, Nashville, TN 37235}
\altaffiliation{Neurodiversity Inspired Science and Engineering Graduate Fellowship}
 
\author[0000-0003-0509-2656]{Matthew Shetrone}
\affiliation{University of California Observatories, University of California Santa Cruz, Santa Cruz, CA, 95064, USA}

\author[0009-0008-0081-764X]{Alessa Ibrahim Wiggins}
\affiliation{Department of Physics and Astronomy, Texas Christian University, TCU Box 298840 \\
    Fort Worth, TX 76129, USA }

\author{John Donor}
\affiliation{Department of Physics and Astronomy, Texas Christian University, TCU Box 298840 \\
Fort Worth, TX 76129, USA }

\author[0000-0002-0740-8346]{Peter M. Frinchaboy}
\affiliation{Department of Physics and Astronomy, Texas Christian University, TCU Box 298840 \\
Fort Worth, TX 76129, USA }
\affiliation{Maunakea Spectroscopic Explorer, Canada-France-Hawaii-Telescope, Kamuela, HI 96743, USA}

\begin{abstract}

In the coming years, detailed chemical abundances from large-scale high-resolution spectroscopic surveys will become available for vast numbers of stars across the Milky Way. Previous work has suggested that abundance ratios from these spectra can allow us to estimate ages from a large number of stars. These data will be leveraged to calibrate chemical clocks to age-date field stars, as reliable stellar ages remain elusive. 
In this work, we extended our empirical relationship between stellar age and their carbon-to-nitrogen ([C/N]) abundance ratio for evolved stars to older and more metal-poor stars   by combining the original open cluster calibration sample and four globular clusters: 47 Tuc, M 71, M 4, and M 5. 
With this extension, [C/N] can be used as a chemical clock for evolved field stars to investigate not only regions within the metal rich disk, but also more metal-poor regions of our Galaxy. 
We have established the [C/N]-age relationship for APOGEE DR17 red giant stars, that have experienced the first dredge up but have not yet undergone any extra-mixing, in clusters 
usable for ages between $8.62 \leq \log(Age[{\rm yr}]) \leq 10.13$ and for metallicites of $-1.2\leq[Fe/H]\leq+0.3$. This relationship can be uniformly applied to these stars within the APOGEE DR17 sample.
This measured [C/N]-age APOGEE DR17 relationship is also shown to be
consistent with stellar ages derived from asterosiesmic results of APOKASC and APO-K2.

\end{abstract}


\section{Introduction} \label{sec:intro}

One of the key uncertainties in mapping the Galaxy's evolution is the ability to accurately measure ages of individual stars. 
While star clusters have long been the gold standard to measure ages, new methods
have begun to allow us to probe the ages of field stars. 
The common method to determine field giant star stellar ages is to compare key parameters, 
whether directly observed (e.g., colors and apparent magnitudes, the latter combined with parallaxes), or 
less directly inferred (e.g., $\log{g}$, $T_{\rm eff}$), to those predicted from stellar evolution models. While this strategy allows for accurate relative aging of stars, it is complicated due to degeneracies in these parameters caused by chemical differences. 
Fortunately, several large-scale spectroscopic surveys are systematically collecting high-resolution data, such as {the Apache Point Observatory Galactic Evolution Experiment survey \citep[APOGEE,][]{dr17}, the Gaia-ESO public spectroscopic survey \citep{gaia_eso}, and the GALactic Archaeology with HERMES survey \citep[GALAH,][]{galah}}. These surveys are providing detailed and precise chemical abundances for multiple elements spanning millions of stars across various Galactic stellar populations, including star clusters.  Apart from providing essential information to break the age-dating degeneracies, these spectroscopic surveys provide additional
key parameters, such as Li-depletion \citep[e.g.,][and references therein]{galahli,lithium}, [s-process/$\alpha$] \citep[e.g.,][]{Casali20_sprocess}, and [C/N] abundance ratios \citep[e.g.,][]{CN_M67,CN_sten,casali_2019}, that are alternative, chemistry-based diagnostics of stellar age.

A key ``chemical clock" that will be available for 100,000s of stars is the measurement of the carbon-to-nitrogen ratio [C/N] in red giant branch (RGB) stars.
{Carbon and nitrogen abundances in evolved stars on the RGB will change due to the first dredge up (FDU), where material from regions that have previously undergone nuclear burning are brought to the surface. The changes in the balance of carbon and nitrogen at the surface are dependant on stellar mass \citep{Masseron_Gilmore_2015, Martig_2016, Ness_2016}. With open cluster studies, we can therefore empirically calibrate a relationship between stellar ages, exploiting the precise ages determined by cluster main sequence turn off (MSTO) isochrone-fitting, and precise [C/N] abundances measured for the RGB stars in these systems \citep{casali_2019,occam_p7}. }

Just obtaining [C/N] from models is not good enough, and so calibrating [C/N]-based ages to other age determinations (e.g., star clusters, asteroseismology) is essential to making it a reliable method.  Asteroseismology, now being generated for large numbers of stars through space missions like CoRoT \citep{corot}, {\em Kepler} \citep{kepler}, and TESS \citep{TESS}, uses high-precision photometry to measure stellar oscillation modes that depend on stellar structure, and therefore stellar masses and radii. With known masses, radii, and compositions, ages for RGB stars can then be precisely inferred from stellar models. For example, utilizing combined data from {\em Kepler} and the APOGEE survey, the APOKASC catalog \citep[][]{APOKASC1,apokasc,APOKASC3} has measured masses, and thereby ages, for $\sim 7000$ evolved stars. 
Such masses and ages have been shown previously to correlate with [C/N] \citep{Martig_2016, Ness_2016}. 
However, obtaining their exact absolute calibration is challenging \citep{Gualme_2016, Huber_2017, Zinn_2019}, as is controlling for physical (e.g. mass loss) and systematic differences. Thus, searching for independent strategies for [C/N]$-age$ calibration is worthwhile.

{With the availability of much more APOGEE data via SDSS DR17, we extend our empirical calibration relationship between [C/N] abundances of RGB stars \citep{occam_p7} to enable age-dating for more {metal-poor} stars {($[Fe/H]<-0.5$)}.
Using the ages determined by \citet{cg20} and \citet{HST_GC_Ages} for clusters that are in
APOGEE/SDSS DR17 as defined by the Open Cluster Chemical Abundances and Mapping (OCCAM) Survey} \citep{occam_p6} and \citet{VB2021_GCmems}, we calibrate a relationship between [C/N] and stellar age for stars that have experienced the first dredge-up and no extra mixing. 
{We also 
compare our results to those from previous work, in particular from the APOKASC and APO-K2 astroseismology methods for determining stellar ages.  These improvements in the [C/N]-age calibration will help increase the power of large-scale, high resolution stellar spectroscopic surveys like APOGEE to explore Galactic chemodynamical evolution using a reliable means to benchmark ages for vast numbers of stars. }

\section{Data}

\subsection{SDSS/APOGEE Survey} \label{sdss_data}

Data used in this study were taken as part of the Apache Point Observatory Galactic Evolution Experiment \citep[APOGEE 1 \& 2;][Majewski et al., {\it in prep}]{apogee}, under the umbrella of the Sloan Digital Sky Survey III \& IV \citep[][respectively]{sdss3,sdss4} . Two spectrographs, APOGEE-North and -South \citep{apogee_inst}, are mounted on 2.5-m telescopes for data collection. The APOGEE-North spectrograph is located on the Sloan Foundation Telescope \citep{sloan_telescope} at the Apache Point Observatory, and the APOGEE-South spectrograph is on the du Pont telescope \citep{du_pont} at Las Campanas Observatory. Observation target selection for the APOGEE surveys is described in \citet{zasowski13,zasowski17}, \citet{ap2n_target}, and \citet{ap2s_target}. Further details regarding the open cluster targeting can be found in \citet{frinchaboy_13} and \citet{donor_18}. \\
\indent The APOGEE data reduction pipeline described in \citet{nidever_2015} was used to reduce the spectra observed during the APOGEE surveys. Key parameters were derived via the APOGEE Stellar Parameters and Chemical Abundance Pipeline \citep[ASPCAP;][]{aspcap}, a fully automated process. The spectral analysis process starts by calculating large grids of synthetic spectra \citep{zamora_15} using stellar atmospheric models \citep{meszaros_2012} and a line list updated with astrophysically informed atomic and molecular data (\citealt{smith21}; see also \citealt{shetrone_2015}). 

The final APOGEE-2 data release, \citep[DR17;][]{dr17}, has $\sim$734,000 stars and includes {all observations from both APOGEE spectrographs taken from August 2011 to January 2021}.
A full description of the APOGEE DR17 data quality and parameter limitations are presented in \citet{dr17}. 
DR17 includes chemical abundances for over 20 elements along with stellar parameters as determined by ASPCAP are reported, including carbon and nitrogen abundances. {Both carbon and nitrogen can be challenging to measure using optical spectra. Still, an infrared survey like APOGEE has the necessary wavelength coverage to reliably measure C, N, and O abundances by combining the analysis of their contributions to the myriad of molecular bands of CN, CO, and OH.} These molecular lines are vital to this work and are fully described in \citet{smith21}.

\subsection{Gaia (E)DR3}

{\em{Gaia}} is a space-based all-sky astrometric and photometric mission from the European Space Agency ({\em ESA}) that probes extensive regions of the Galactic disk. Hundreds of star clusters were cataloged \citep{gaia_dr2, cg_18} 
after the second data release of Gaia, and both membership and fundamental parameter estimates (distance, age) were improved for over 1800 clusters observed \citep{cg20}.
Gaia \citep[EDR3;][]{gaia_edr3} was released on 3 December 2020. Gaia Data Release 3 (Gaia DR3) was released on 13 June 2022. We used \citet{VB2021_GCmems} as a part of our membership method which uses mean parallaxes and proper motions from Gaia EDR3, which are unchanged in DR3.

\subsection{Globular Cluster Membership \label{GC_Mems}}
To select stars with reliable carbon and nitrogen measurements
for our analysis, we followed quality cuts outlined in Table 1 of \citet{occam_p7} for bitwise flags, and signal-to-noise ratio flag (SNREV).
{We analyzed the full APOGEE-2 DR17 globular cluster sample \citet{Schiavon_GCVAC} to determine for which clusters this analysis could work, as we need to isolate first population stars without extra enrichment or mixing, which requires reach to near the base of the RGB.}
Globular clusters used in this work were selected because (1) they were known clusters that were observed in the APOGEE survey as well as (2) have reliable ages in \citet{HST_GC_Ages}, and (3) have observed stars in the required range of $\log(g)$, lower giant brach needed to be able to isolate first population stars without extra mixing.
Reliable cluster members were determined using the APOGEE membership flag (MEMBER) and \citet{VB2021_GCmems} Gaia based membership probabilities that were greater than 0.999. We compared this membership method with methods within \citet{occam_p6} and \citet{Schiavon_GCVAC}, and found all three methods obtained the same stellar membership for all globular clusters used; therefore, combining the \citet{occam_p7} sample with our globular cluster sample does not lose membership uniformity. 
{We find only a limited sample of globular cluster were observed by APOGEE to the depth needed to enable our analysis, described in \S\ref{sec:isolate}, which are presented in Table \ref{sum_table}.   Other APOGEE globular clusters, for which preliminary work was done, are not presented in this work.}

\begin{deluxetable}{lccc}[h!]
 \tablecaption{Globular Cluster Sample\label{sum_table}}
 \tablewidth{0pt}
 \tabletypesize{\small}
 \tablehead{
 \colhead{Cluster}& 
 \colhead{$\log$(Age)\tablenotemark{a}} &  
 \colhead{ [Fe/H]\tablenotemark{b} } &
 \colhead{ [C/N]\tablenotemark{b} } \\[-1.5ex]
 \colhead{name} &
 \colhead{(yr)} &
 \colhead{(dex)} & 
 \colhead{(dex)} 
 }
\startdata
 47 Tuc  &  10.13  &  $-$0.72  &  $+0.16 \pm 0.04$  \\
 M 71    &  10.13  &  $-$0.78  &  $+0.09 \pm 0.12$  \\
 M 4     &  10.11  &  $-$1.04  &  $+0.00 \pm 0.31$  \\
 M 5     &  10.08  &  $-$1.20  &  $-0.13 \pm 0.22$  
\enddata
\tablenotetext{a}{All ages come from \citet{HST_GC_Ages} who report a goodness of fit for {their relative ages, not an age uncertainty, therefore we adopt a conservative $\pm0.1$ $\log$(Age) uncertainty, that is larger than their upper and lower limits on their $\log$(Age), to account for any systemics, and is consistent with the \citet{cg20} ages uncertainties.}}
\tablenotetext{b}{{Cluster [Fe/H] and [C/N] measurements determined using only identified the first generation stars, whose identification is described in \S\ref{sec:isolate}, with those stars presented in Table \ref{tab:GC_stars}.}}
\end{deluxetable}
\begin{deluxetable*}{llrrrrr}[t!]
    \tablecaption{Calibration Globular Cluster Sample Stellar Data from APOGEE DR17. \label{tab:GC_stars}} 
    \tablewidth{0pt}
    \tabletypesize{\normalsize}
    \tablehead{
    \colhead{Cluster} & 
    \colhead{2MASS ID} &
    \colhead{$T_{\rm eff}$} &
    \colhead{$\log(g)$} &
    \colhead{[Fe/H]} & 
    \colhead{[C/N]} \\[-0.5ex] 
    \colhead{Name} &
    \colhead{} &
    \colhead{(K)} &
    \colhead{(dex)} & 
    \colhead{(dex)} & 
    \colhead{(dex)} 
    }
    \startdata
    47 Tuc & 2M00204027$-$7201425 & $4753 \pm 12$ & $2.46 \pm 0.04$ & $-0.69 \pm 0.01$ & $+0.16 \pm 0.13$\\
    47 Tuc & 2M00215813$-$7158147 & $4809 \pm 15$ & $2.46 \pm 0.04$ & $-0.71 \pm 0.01$ & $+0.16 \pm 0.11$\\
    47 Tuc & 2M00231815$-$7211516 & $4814 \pm 16$ & $2.39 \pm 0.04$ & $-0.76 \pm 0.01$ & $+0.13 \pm 0.15$\\
    47 Tuc & 2M00251382$-$7159103 & $4813 \pm 15$ & $2.37 \pm 0.04$ & $-0.79 \pm 0.01$ & $+0.17 \pm 0.12$\\
    47 Tuc & 2M00233065$-$7150017 & $4756 \pm 12$ & $2.36 \pm 0.03$ & $-0.70 \pm 0.01$ & $+0.16 \pm 0.11$\\
    47 Tuc & 2M00235608$-$7141488 & $4811 \pm 15$ & $2.39 \pm 0.04$ & $-0.75 \pm 0.01$ & $+0.25 \pm 0.20$\\ \hline
    M 71   & 2M19533271$+$1841101 & $4825 \pm 15$ & $2.73 \pm 0.04$ & $-0.70 \pm 0.01$ & $+0.10 \pm 0.07$\\ 
    M 71   & 2M19533114$+$1845204 & $4858 \pm 16$ & $2.63 \pm 0.04$ & $-0.71 \pm 0.01$ & $+0.09 \pm 0.07$\\ 
    M 71   & 2M19533428$+$1846550 & $4854 \pm 11$ & $2.47 \pm 0.03$ & $-0.68 \pm 0.01$ & $+0.18 \pm 0.13$\\ 
    M 71   & 2M19533989$+$1844229 & $4858 \pm 10$ & $2.45 \pm 0.03$ & $-0.72 \pm 0.01$ & $+0.22 \pm 0.16$\\ 
    M 71   & 2M19534992$+$1841255 & $4914 \pm 19$ & $2.94 \pm 0.04$ & $-0.71 \pm 0.01$ & $+0.13 \pm 0.09$\\ 
    M 71   & 2M19535769$+$1844567 & $4866 \pm 16$ & $2.61 \pm 0.04$ & $-0.68 \pm 0.01$ & $-0.22 \pm 0.20$\\ 
    M 71   & 2M19535018$+$1845525 & $4751 \pm 13$ & $2.38 \pm 0.04$ & $-0.73 \pm 0.01$ & $+0.01 \pm 0.03$\\ 
    M 71   & 2M19540228$+$1842447 & $4772 \pm 15$ & $2.37 \pm 0.04$ & $-0.76 \pm 0.01$ & $+0.07 \pm 0.05$\\ 
    M 71   & 2M19533593$+$1847564 & $4757 \pm 10$ & $2.40 \pm 0.03$ & $-0.75 \pm 0.01$ & $-0.04 \pm 0.12$\\ \hline
    M 4    & 2M16223348$-$2631308 & $5050 \pm 13$ & $2.82 \pm 0.04$ & $-1.08 \pm 0.01$ & $+0.37 \pm 0.26$\\
    M 4    & 2M16225050$-$2642162 & $5291 \pm 16$ & $2.28 \pm 0.04$ & $-1.22 \pm 0.01$ & $+0.00 \pm 0.18$\\
    M 4    & 2M16231475$-$2645281 & $5063 \pm 14$ & $2.98 \pm 0.04$ & $-1.05 \pm 0.01$ & $+0.51 \pm 0.37$\\
    M 4    & 2M16232148$-$2638354 & $5272 \pm 16$ & $2.23 \pm 0.04$ & $-1.20 \pm 0.01$ & $-0.37 \pm 0.27$\\
    M 4    & 2M16233193$-$2631314 & $5397 \pm 39$ & $2.23 \pm 0.06$ & $-1.22 \pm 0.02$ & $-0.25 \pm 0.47$\\
    M 4    & 2M16233236$-$2629222 & $4932 \pm 12$ & $2.62 \pm 0.04$ & $-1.05 \pm 0.01$ & $+0.18 \pm 0.18$\\
    M 4    & 2M16233621$-$2640002 & $4982 \pm 17$ & $2.62 \pm 0.04$ & $-1.08 \pm 0.01$ & $-0.22 \pm 0.22$\\ \hline
    M 5    & 2M15175206$+$0159462 & $4994 \pm 14$ & $2.45 \pm 0.04$ & $-1.21 \pm 0.01$ & $+0.22 \pm 0.26$\\
    M 5    & 2M15175554$+$0217164 & $4991 \pm 17$ & $2.35 \pm 0.04$ & $-1.12 \pm 0.01$ & $-0.13 \pm 0.10$\\
    M 5    & 2M15181619$+$0205358 & $5001 \pm 14$ & $2.29 \pm 0.04$ & $-1.15 \pm 0.01$ & $-0.28 \pm 0.23$\\
    M 5    & 2M15182846$+$0159283 & $4866 \pm 16$ & $2.22 \pm 0.05$ & $-1.29 \pm 0.01$ & $+0.16 \pm 0.22$\\
    M 5    & 2M15182917$+$0159269 & $4908 \pm 17$ & $2.17 \pm 0.05$ & $-1.23 \pm 0.01$ & $-0.45 \pm 0.33$\\
    M 5    & 2M15183720$+$0208197 & $4887 \pm 29$ & $2.16 \pm 0.07$ & $-1.26 \pm 0.02$ & $-0.24 \pm 0.21$\\
    M 5    & 2M15183873$+$0208200 & $4890 \pm 14$ & $2.20 \pm 0.04$ & $-1.16 \pm 0.01$ & $-0.13 \pm 0.09$\\
    M 5    & 2M15183915$+$0205301 & $4940 \pm 19$ & $2.37 \pm 0.05$ & $-1.23 \pm 0.01$ & $+0.17 \pm 0.13$\\
    M 5    & 2M15183975$+$0212333 & $4992 \pm 19$ & $2.39 \pm 0.05$ & $-1.23 \pm 0.01$ & $-0.24 \pm 0.35$\\ \hline
    \enddata
    \tablecomments{This only includes DR17 stars isolated for the [C/N] calibration relation described in \S\ref{sec:isolate}.}
\end{deluxetable*}
\section{Analysis}

To obtain a more complete picture of the older part of the Milky Way (e.g., bulge, streams, and halo) through the chemical clock [C/N], we need to expand our calibration sample to include the older and more metal poor stars where good ages can be derived, i.e., 
globular clusters. Since globular clusters are among the oldest objects in our universe, they provide a snapshot into the early stages of our Milky Way. 
However, moving forward with utilizing globular clusters in our calibration introduces additional layers of complexity. Firstly, as lower metallicity stars have been observed to have extra mixing effects and second many globular clusters have multiple populations of stars.

In \citet{Shetrone_2019} and \citet{Gratton_2000}, it was found that stars more metal poor than [Fe/H] $ \lesssim -0.5$ will experience extra-mixing along the RGB; this extra-mixing can be significant enough that the link between [C/N] and age may change. To exclude the effects of extra-mixing, additional cuts must be made in surface gravity to select stars that have experienced the FDU but have not undergone any further extra mixing.  This extra mixing happens to occur near the red giant branch bump \citep{Fraser2022, TayarJoyce2022}, and hence, the optimal surface gravity cuts would be right before this evolutionary phase.

\begin{figure*}[t!]
    \centering
    \plottwo{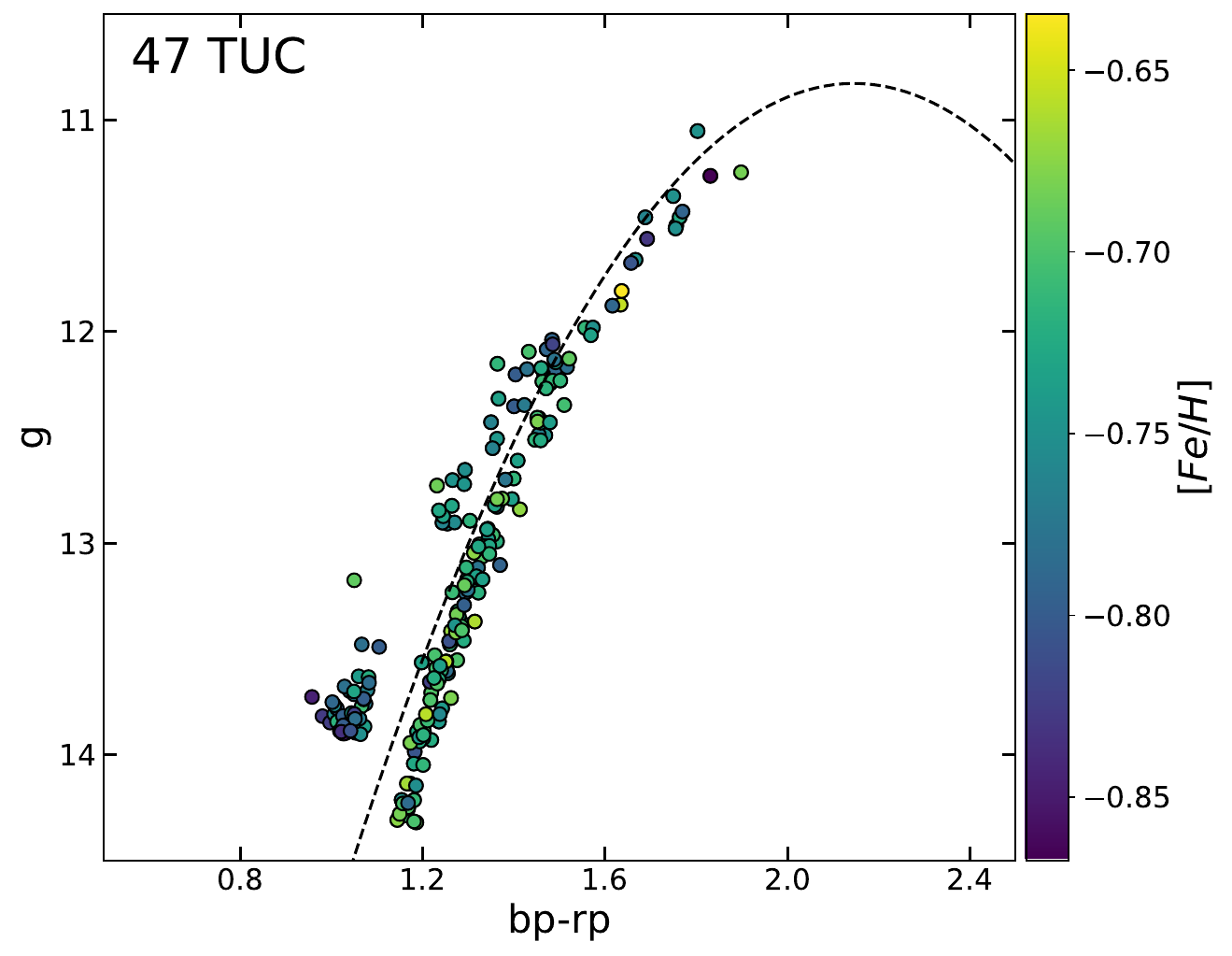}{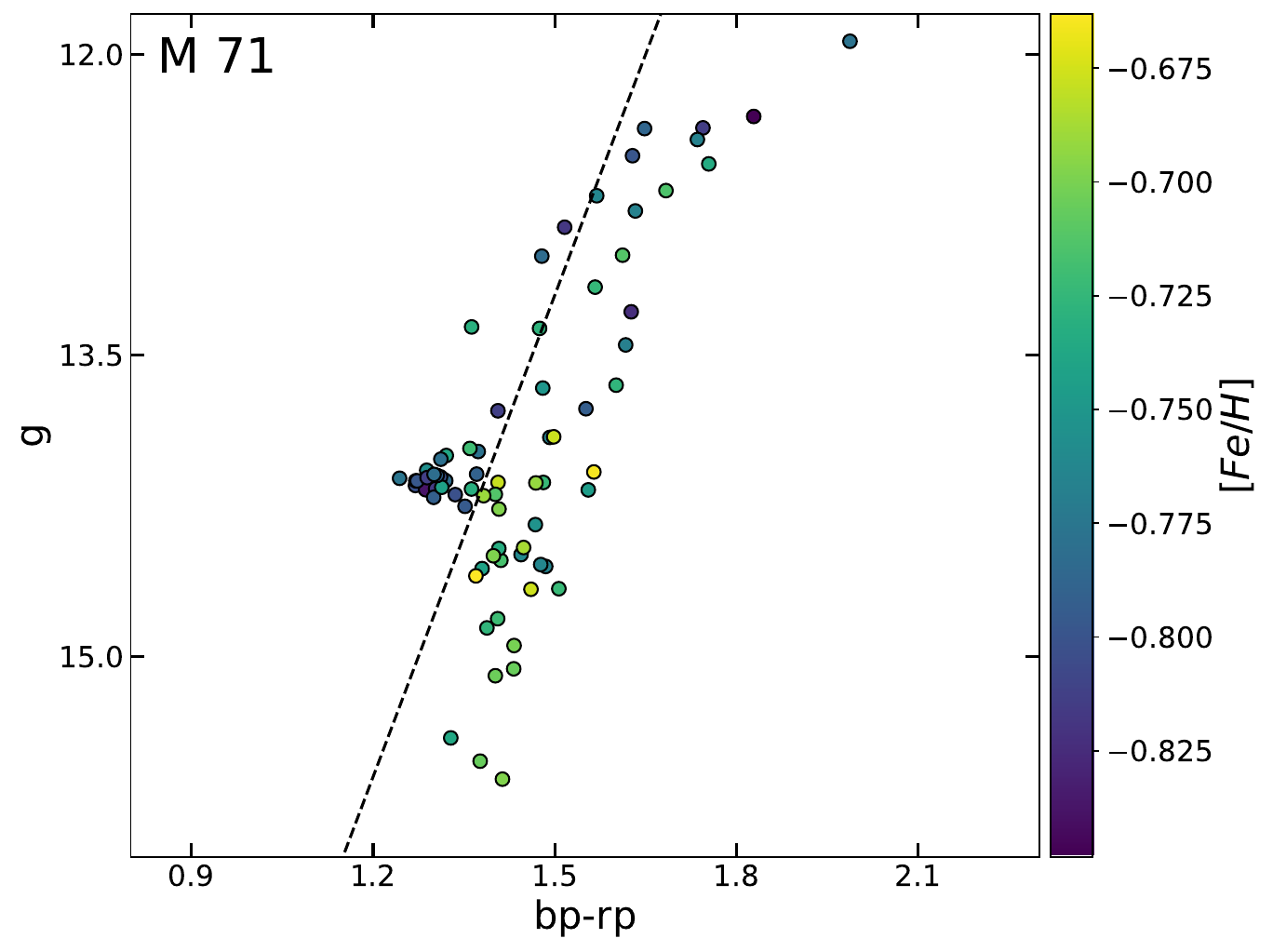}
    \plottwo{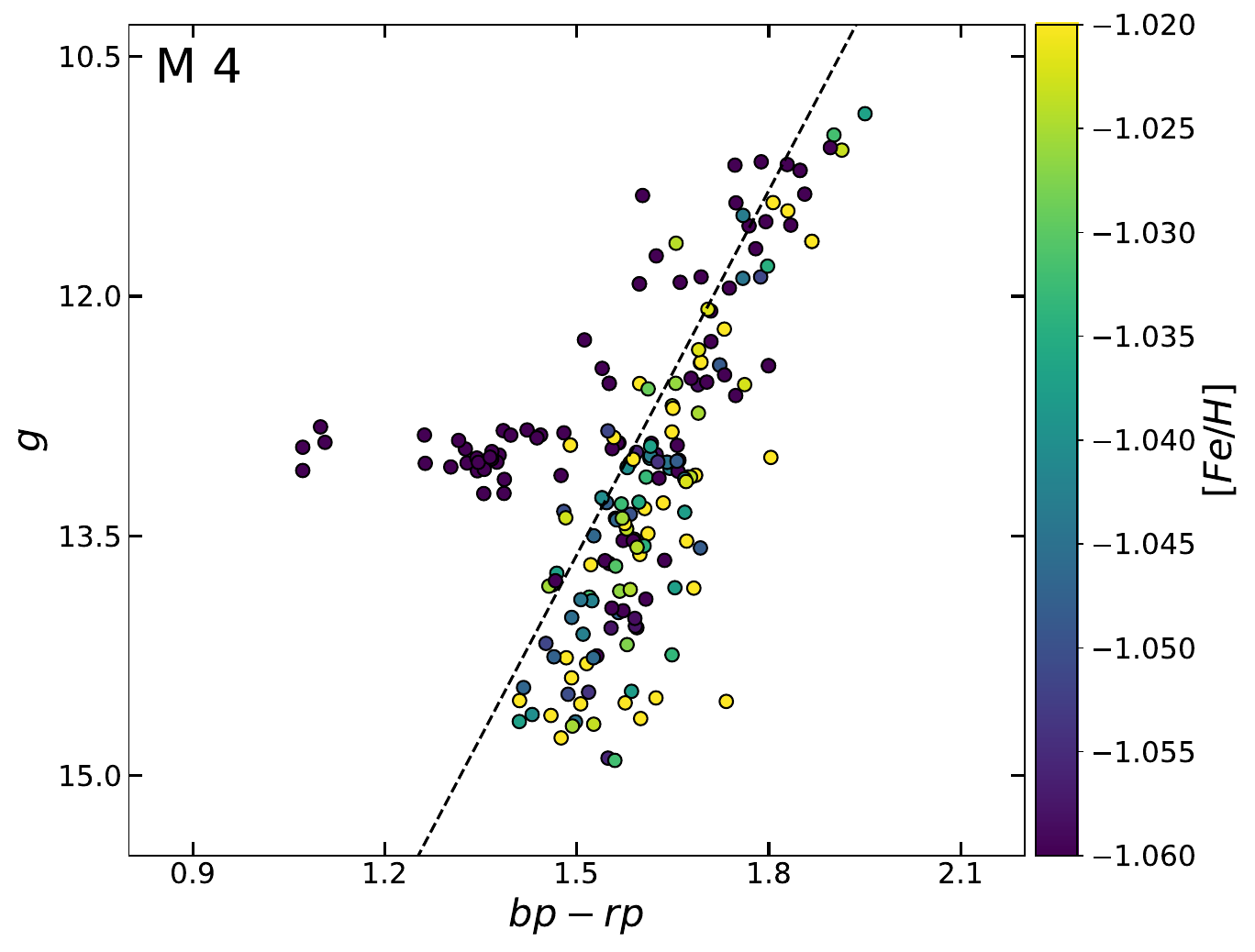}{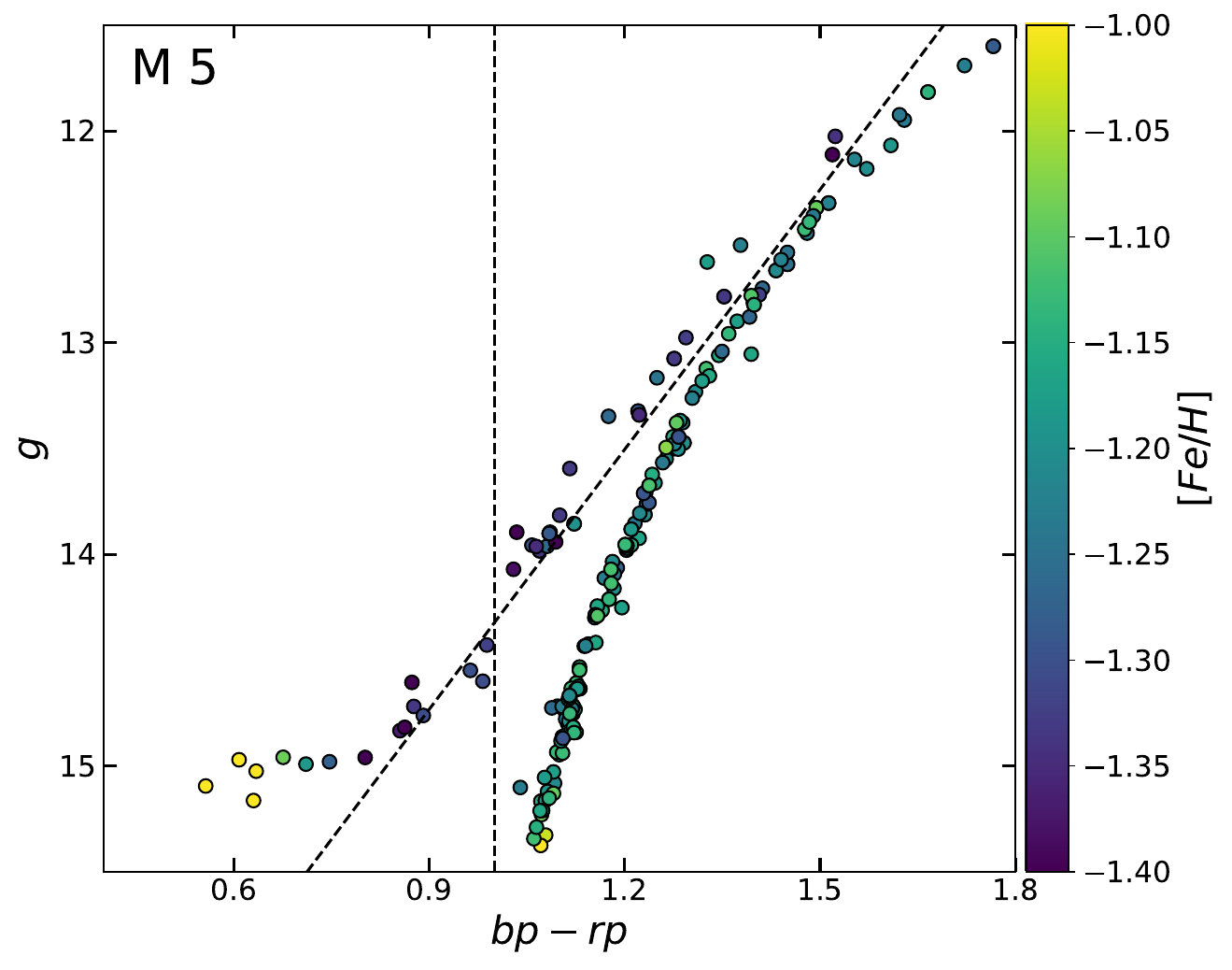}
    \caption{Gaia CMD for globular clusters 47 TUC, M 71, M 4, and M 5 colored by [Fe/H]. Black-dashed lines represent the polynomial functions used to separate RGB from RC and AGB members of each cluster. Note the magnitude errors are smaller than the markers.
    }
    \label{RGB_selection}
\end{figure*}

Unlike open clusters, {most} globular clusters have more than one population of stars
\citep[e.g.,][and references therein]{Piotto09,AnnualReview2022_GCMultiPop}, and therefore we need to be able to separate the two populations. The first population of stars will have similar chemical compositions as the Galactic-field stars with analogous metallicity and can presumably be used to calibrate the [C/N]-age relation. The second population, in contrast,  possess peculiar abundances in some light-elements compared to field stars due to the unique enrichment history of the cluster. The elemental abundance variation is found in both RGB stars and MS stars implying the chemical variation is not due to stellar evolution \citep{AnnualReview2022_GCMultiPop}. 
For light elements such as C, N, O, and Na,
second population stars will be enhanced with N and Na while depleted in C and O when compared to the first population. These elements have {clearly identified 
relationships between abundances,} such as the Na-N and C-O correlations and Na-O and C-N anti-correlations, that are found as a prevalent feature across globular clusters with multiple populations. Another relationship found in a majority of globular clusters is a correlation between [Al/Fe] and [Na/Fe]. 
Since these (anti)correlations exist, we can use them to separate the first population of stars from the second population of stars within their corresponding chemical spaces, and to exclude these stars with {altered} 
chemistry from our age calibration abundance {similar to that of previous work on globular clusters with APOGEE \citep[e.g.,][]{Sz2015,Schiavon_GCVAC,Horta2020}. }

\subsection{Calibration Sample Selection\label{sec:isolate}}

\begin{figure*}[h!]
    \centering
    \plotone{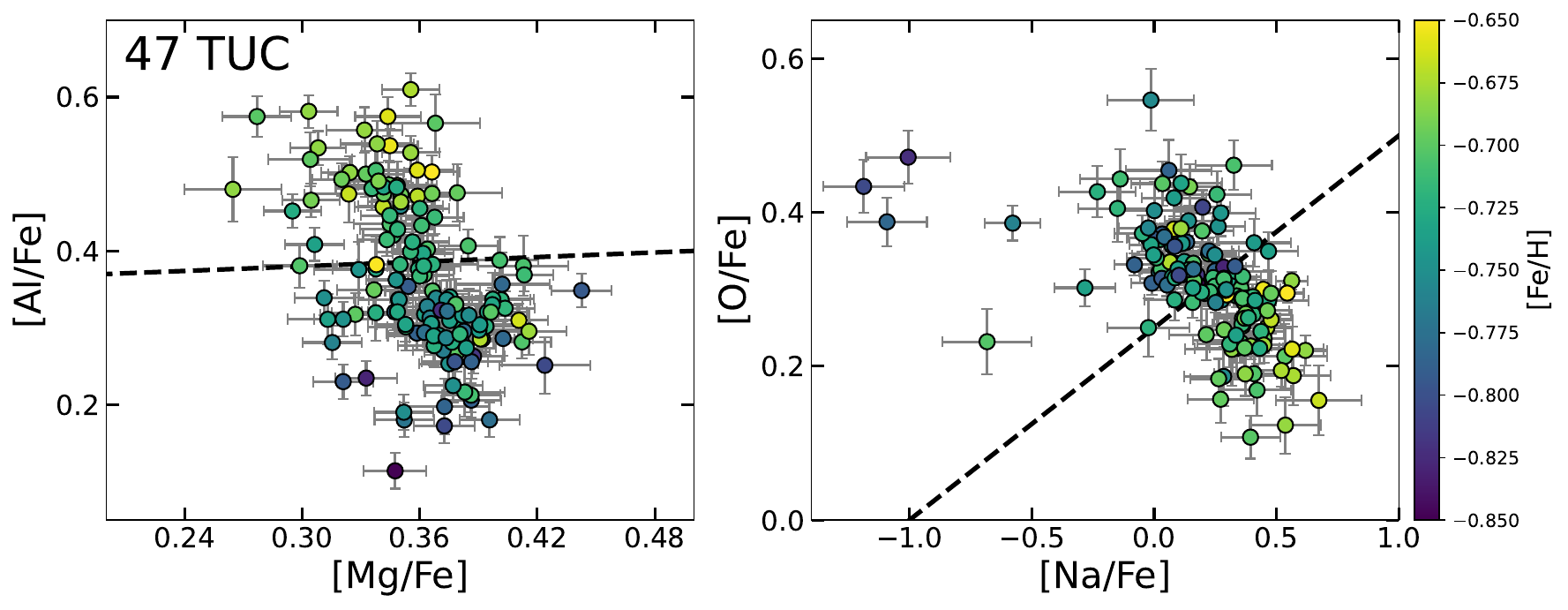}
    \plotone{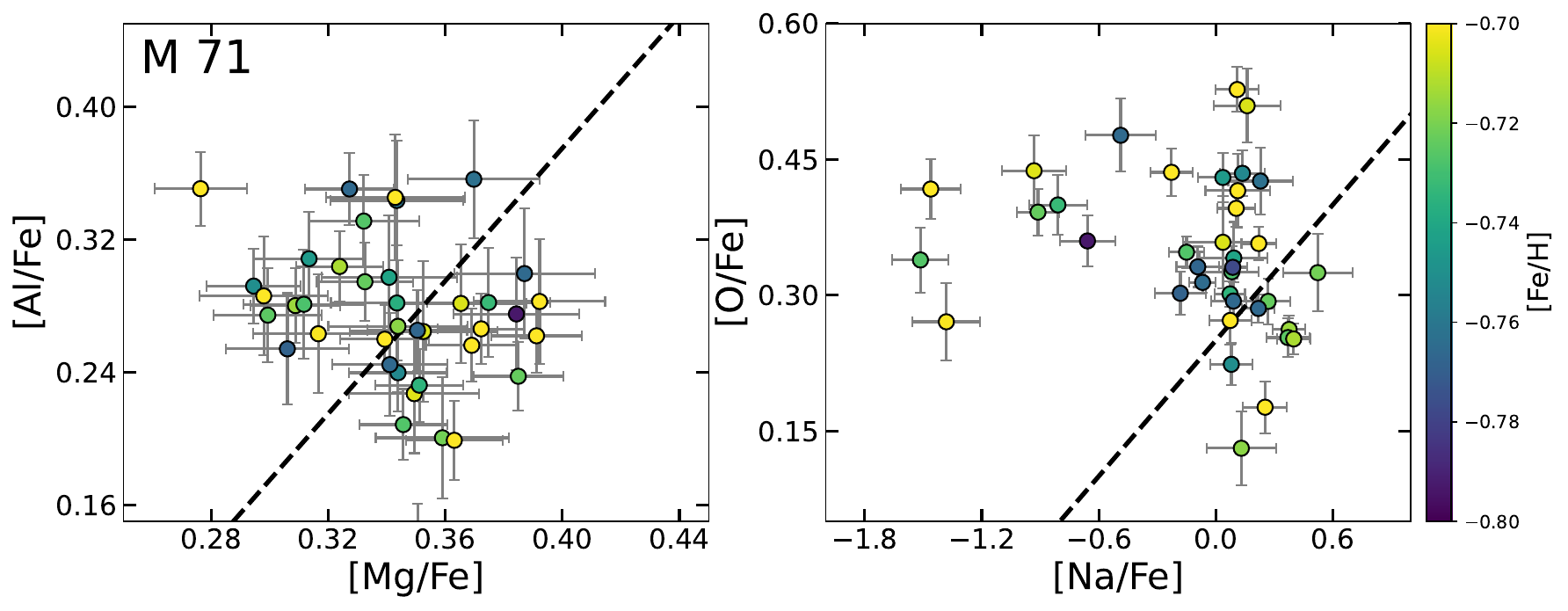}
    \caption{[Al/Fe]-[Mg/Fe] and [O/Fe]-[Na/Fe] (anti)Correlation plots for globular clusters 47 TUC ({\it top set}) and M 71 ({\it bottom set}) colored by [Fe/H]. Black-dashed lines represent functions to isolate the two populations of stars within the RGB sample.}
    \label{Gen1_seletion_rich}
\end{figure*}

\begin{figure*}[h!]
    \centering
    \plotone{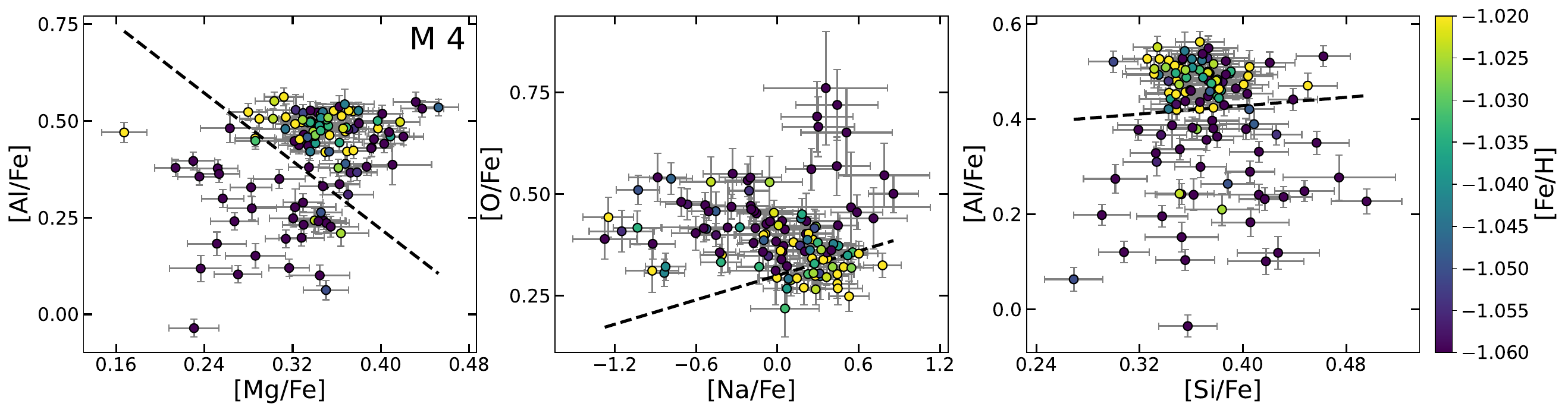}
    \plotone{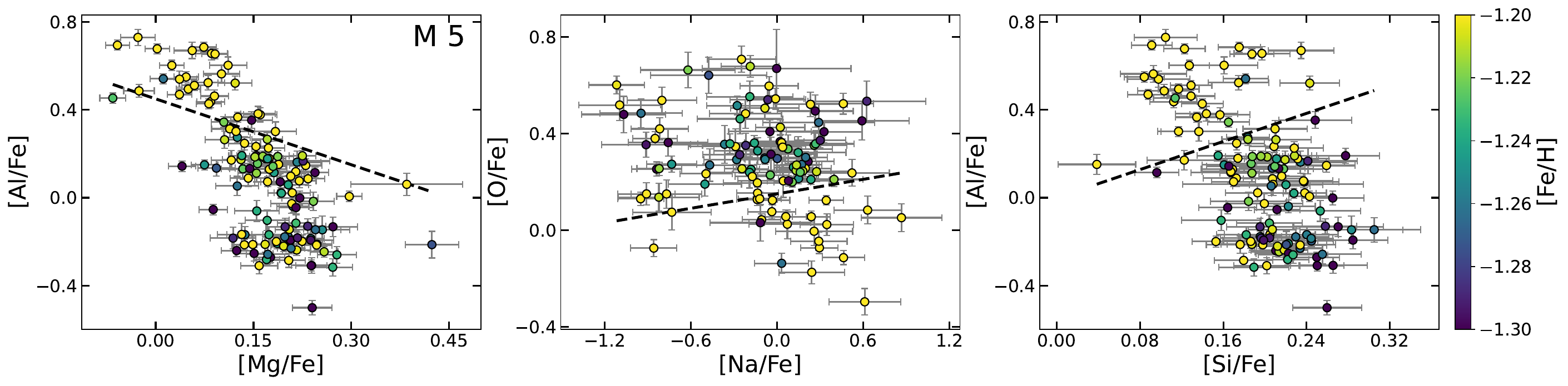}
    \caption{[Al/Fe]-[Mg/Fe], [O/Fe]-[Na/Fe], and [Al/Fe]-[Si/Fe] (anti)Correlation plots for globular clusters M 4 ({\it top set}) and M 5 ({\it bottom set}) colored by [Fe/H]. Black-dashed lines represent functions to isolate the two populations of stars within the RGB sample.}
    \label{Gen1_seletion_poor}
\end{figure*}

For us to select stars that are reliable for our calibration extension, we need to obtain RGB stars from the first stellar population formed in the cluster. First, we selected RGB stars using a polynomial fit within color-magnitude space, both in Gaia and 2MASS photometry, and then applied offsets until it clearly delineates the AGB stars from the first ascent RGB stars, so they could be rejected, as shown in Figure \ref{RGB_selection}. 
Once the RGB stars were isolated, we separated the populations using known (anti)correlations in chemical space, as shown in Figures \ref{Gen1_seletion_rich} and \ref{Gen1_seletion_poor} {to attempt to isolate only first population stars.} 

\begin{figure*}[t!]
    \centering
    \plottwo{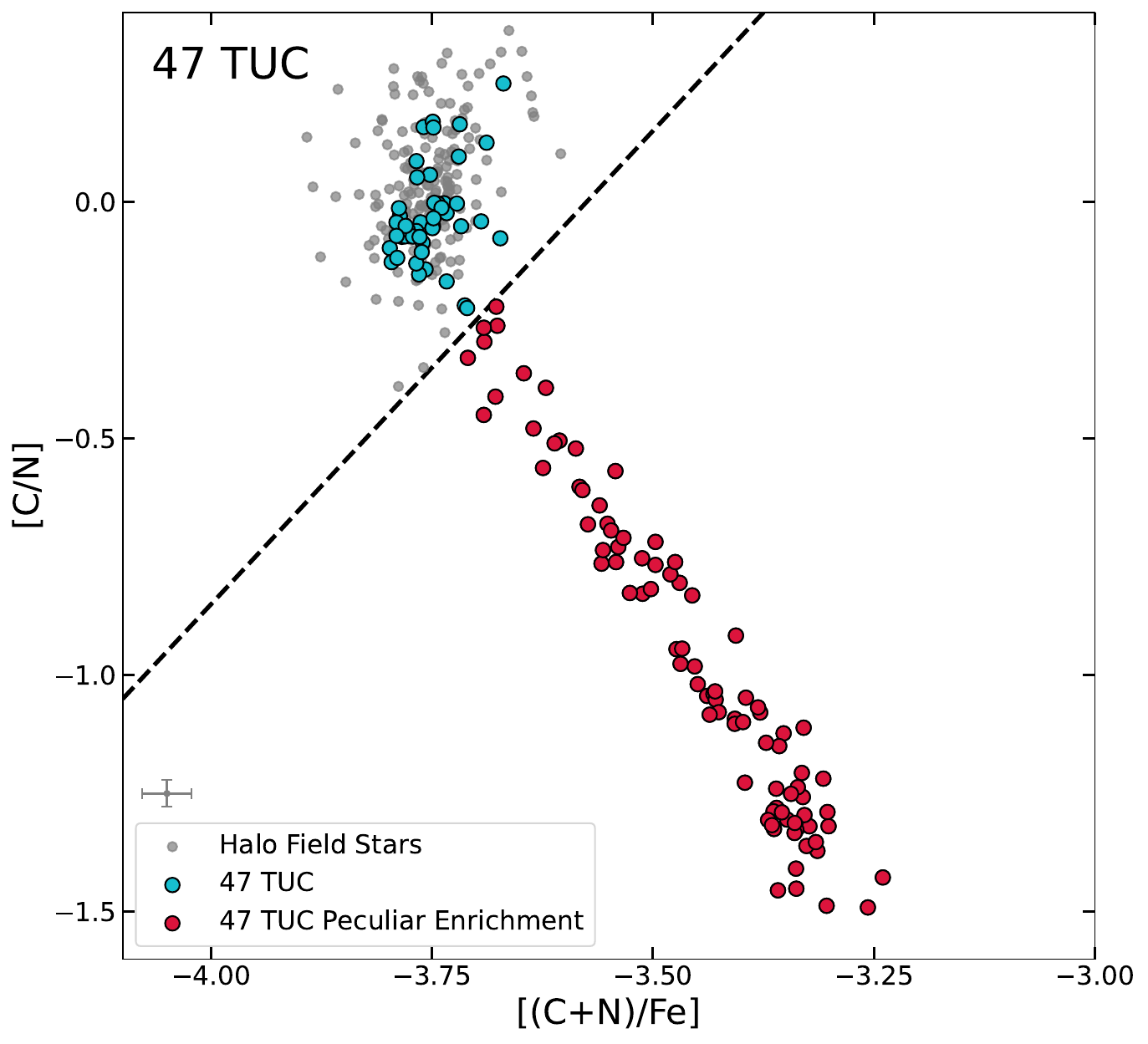}{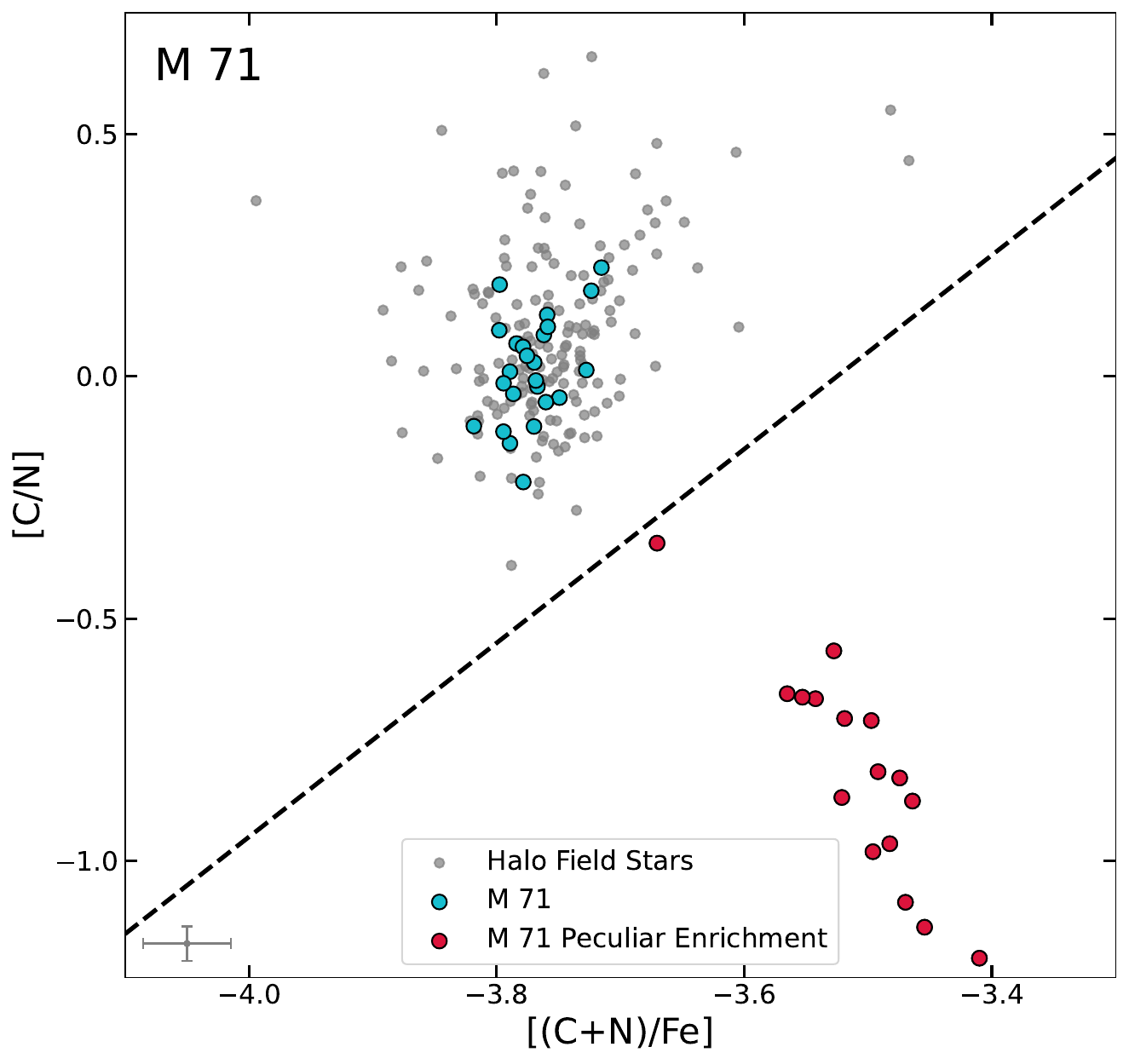}
    \plottwo{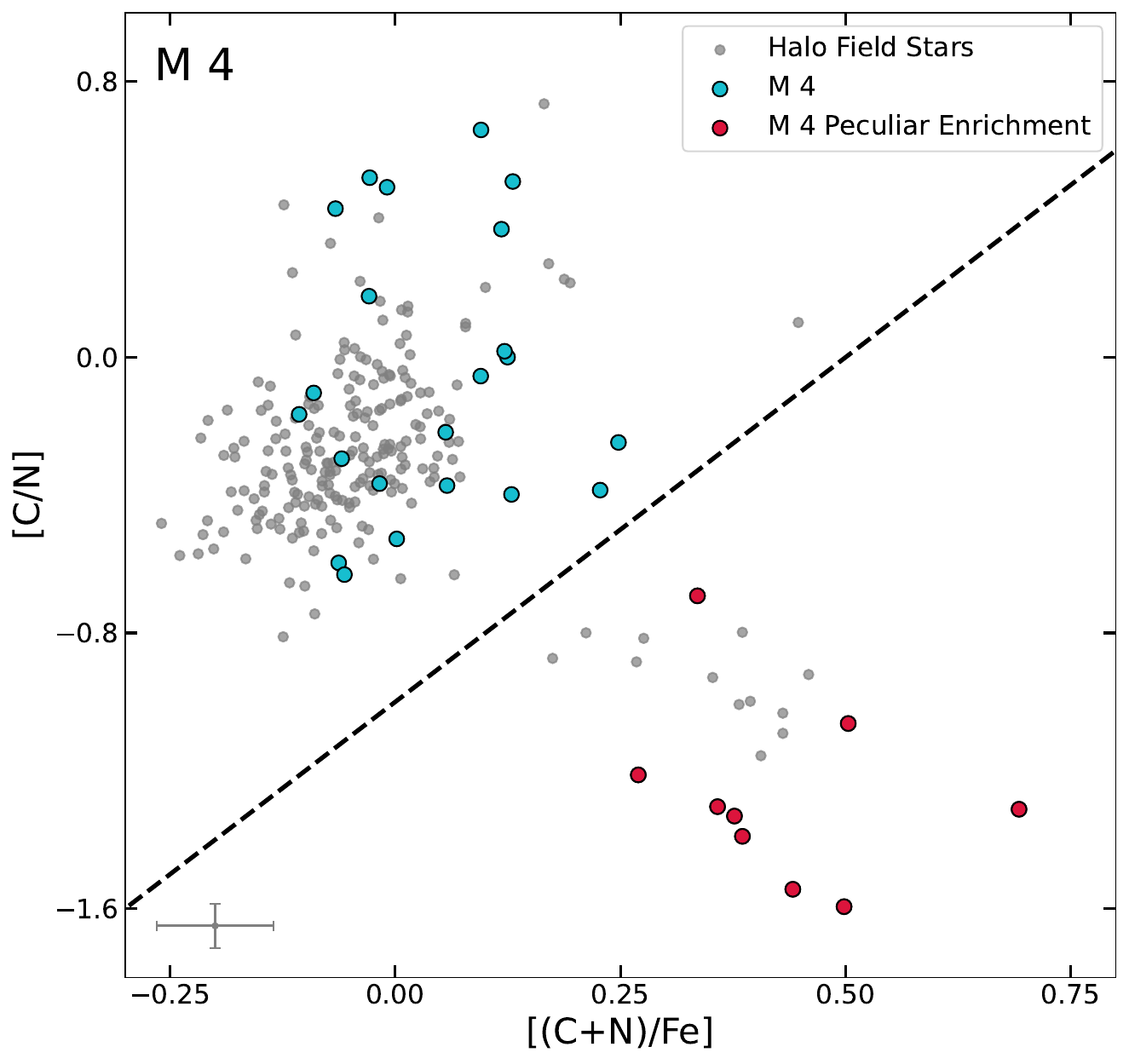}{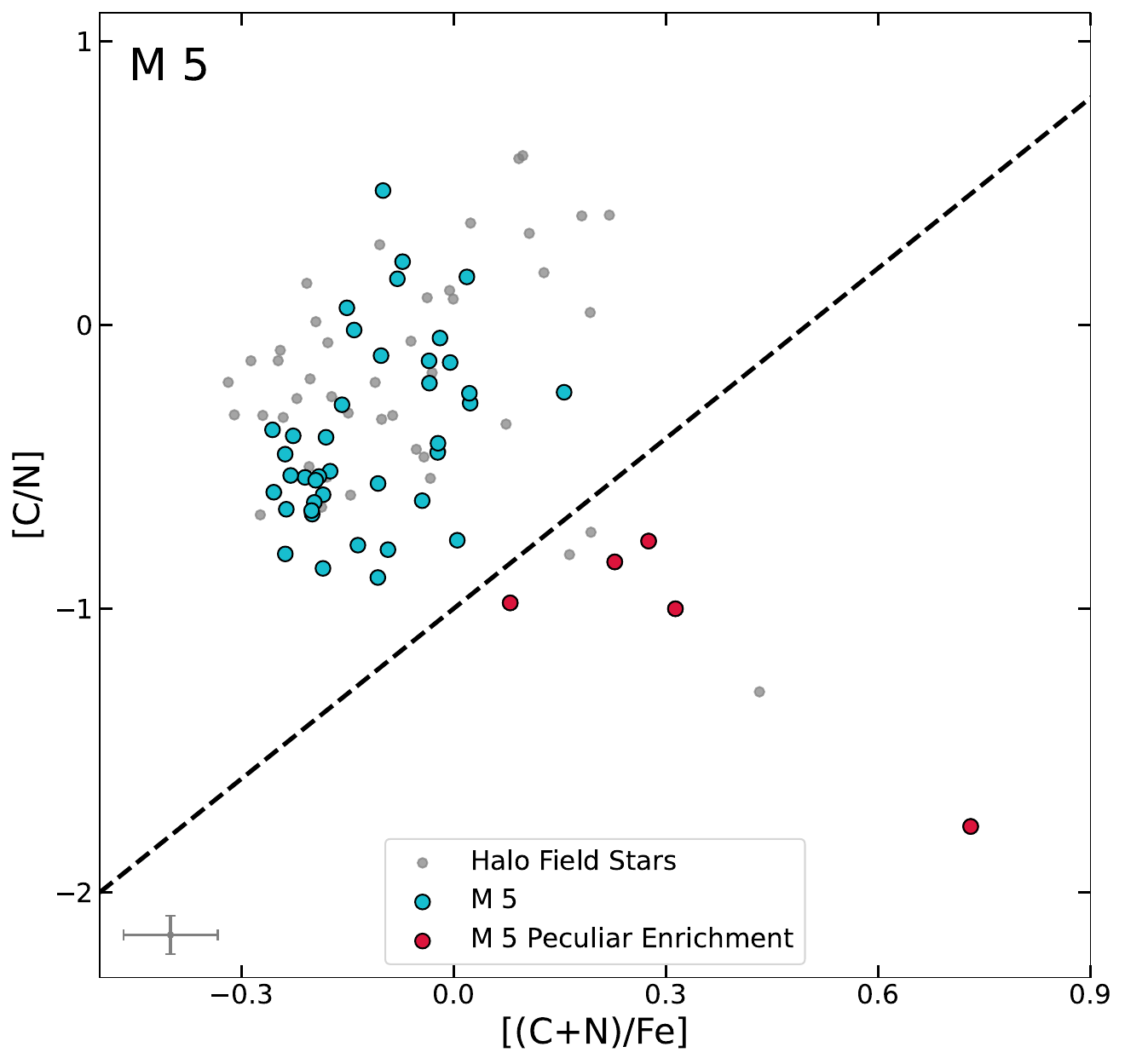}
    \caption{[C/N] versus [C+N/Fe] plots for globular clusters 47 TUC, M 71, M 4, and M 5. Light grey dots are APOGEE halo field stars with similar Mg- and Fe-abundances as the cluster. Stars outlined in black are members of the respective cluster, where those colored in red are considered peculiar enriched stars within the cluster, while those colored in cyan are those that are not peculiar and are similar to the field stars. The black dashed line represents the cut we did to remove the peculiar stars from the calibration sample. A representative error bar is shown in the bottom left corner of each panel for each cluster.}
    \label{CpluN_selection}
\end{figure*}

\begin{figure}
    \centering
    \epsscale{1.17}
    \plotone{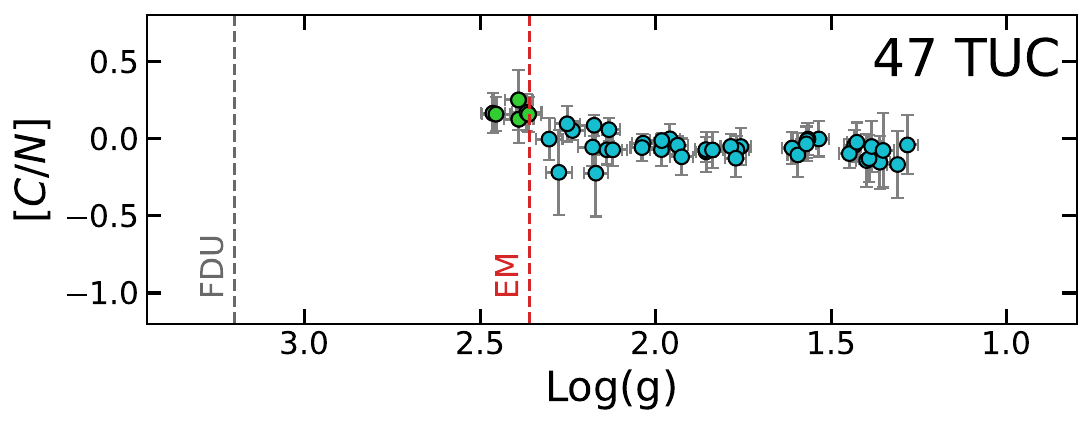}
    \plotone{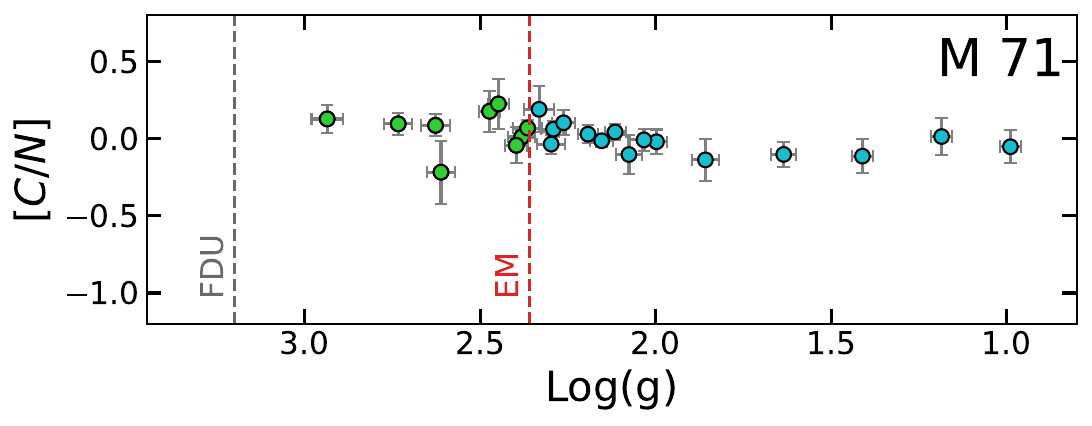}
    \plotone{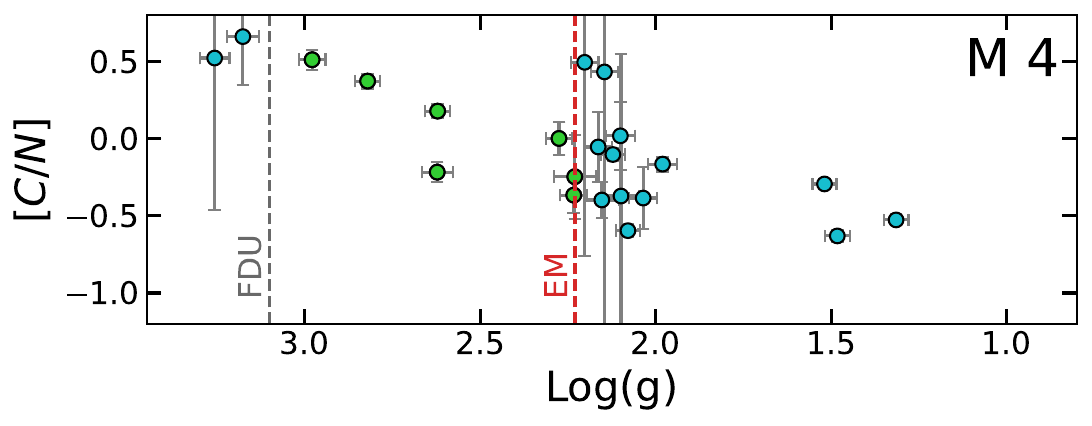}
    \plotone{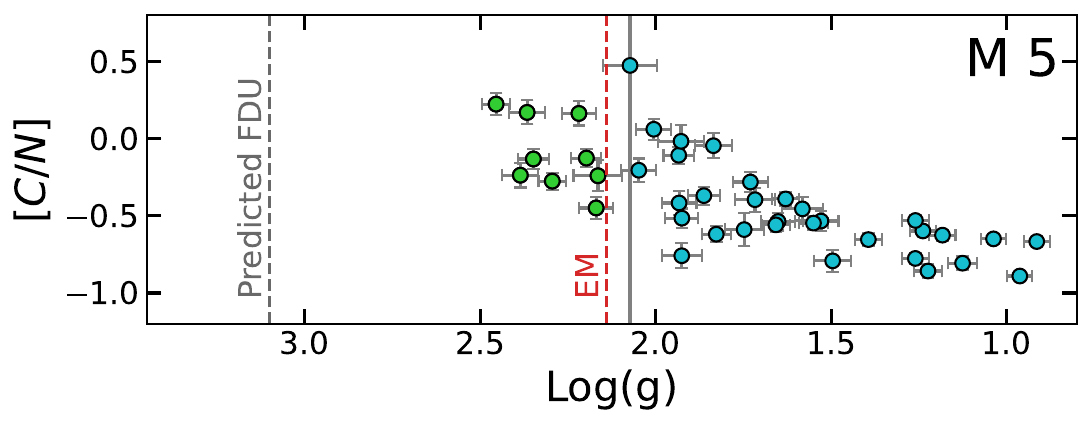}
    \caption{Log(g) vs [C/N] plot of first population of stars without peculiar enrichment. These are the same stars from Figure \ref{CpluN_selection} shown in cyan. Stars used in our calibration are shown in green. The grey vertical dashed line and vertical red dashed line represent the FDU log(g) cut and extra-mixing (EM), respectively, from Table 2 of \citet{Shetrone_2019}.\\}
    \label{loggCuts}
\end{figure}

{
 Next, checked [C/N] versus [C+N/Fe] and we find that our supposed cleaned first population sample still contains additional peculiar enriched stars.  As we only want first population star similar to those found in the Milky Way halo for our calibration, we used the [C/N] versus [C+N/Fe] plots to remove peculiar enriched stars Figure \ref{CpluN_selection}.  The field stars are APOGEE halo stars, selected using the APOGEE targeting flags ({\tt APOGEE2\_TARGET2 = 20 and 21}) to be within scatter of the [Fe/H] and [Mg/Fe] of the cluster, typically within 0.15 dex of the cluster mean, giving us a sample of halo field stars similar [Fe/H] and [Mg/Fe] to the first population selection, to verify that our remaining stars provide a good representation of the field star sample.}
 Following these selections, we applied surface gravity cuts from Table 2 of \citet{Shetrone_2019}, to remove stars that have experienced the FDU but have not undergone extra mixing, shown in Figure \ref{loggCuts}. 
After applying this {criterion}, we show our globular cluster sample summary
in Table \ref{sum_table} which includes: cluster names, cluster ages reported from \citet{HST_GC_Ages}, DR17 average [Fe/H], and DR17 average [C/N] for member stars with reliable measurements that best represent the field sample.

\section{Results}

After application of these additional membership and cluster age cuts the sample used for our [C/N]-age calibration (Figure ~\ref{CN_EXTN_FIT}) is comprised of the 49 clusters (530 stars) from \citet{occam_p7}\footnote{{We note that a few of the outlier clusters from \citet{occam_p7} are single star clusters that may be affected by either poor membership and/or the measured star may be a binary affecting the [C/N] determination.  We keep the full sample \citet{occam_p7} for this work.}} plus four globular clusters: 47 TUC, M 71, M 4, and M 5 (31 stars). The full sample covers an age range of $8.62 \leq \log(Age[{\rm yr}]) \leq 10.13$ and a metallicity range of $-1.2 \leq {\rm [Fe/H]} \leq +0.3$. 
The individual globular cluster stars used in the calibration extension are shown in Table \ref{tab:GC_stars}. 

\subsection{The Extended DR17 [C/N] Abundance/Age Calibration}

In log-log space the relationship between stellar age and [C/N] appears to follow polynomial curve; our best fit is given by:

{\small
\begin{equation} \label{extn_cali_eqn}
\begin{split}
\log{[Age({\rm yr})]}_{\rm DR17} =   -1.78\,(\pm 0.98) \;{\rm [C/N]}^2 + \\ 1.37\,(\pm 0.63)\;{\rm [C/N]} + 10.10\,(\pm 0.10)
\end{split}
\end{equation}}

\begin{figure*}[ht!]
\epsscale{1.15}
\plotone{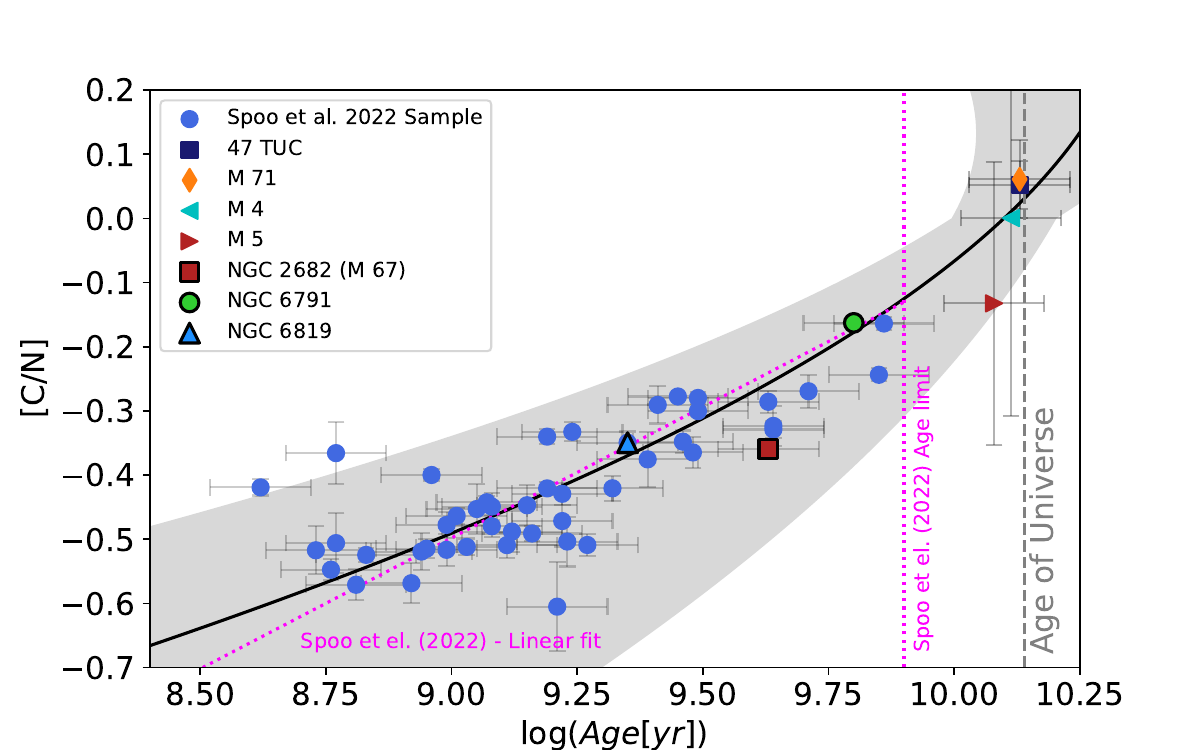}
\centering
\caption{\setlength{\baselineskip}{15.0pt} The [C/N] versus $\log(Age[{\rm yr}])$ distribution for the final sample, composed of \citet{occam_p7} open cluster sample and globular clusters: 47 TUC, M 71, M 4, and M 5. Clusters represented by blue circles are from the final calibration sample of \citet{occam_p7}. Globular clusters 47 TUC, M 71, M 4, and M 5, are represented with an indigo square, an orange thin-diamond, a cyan left-triangle, and a red right-triangle, respectively. Our new fit is shown as a solid black line, and the {maximum} possible calculated age envelope is shown with the grey region. The accepted age of the universe \citep[$13.8 \pm 0.05$ Gyr;][]{age1,Desi1a,Desi1b}
is shown as a vertical grey-dashed line. The open clusters used in our asteroseismic age comparison are NGC 2682 (M 67), NGC 6791, and NGC 6819 and are shown in black outlined red square, green circle, and blue triangle, respectively. NOTE: {Uncertainties on the ages} for the globular clusters are assumed by this work to be $\pm 0.1$ to be conservative and are likely smaller.}
\label{CN_EXTN_FIT}
\end{figure*}

The polynomial fit given by Equation \ref{extn_cali_eqn} and shown in Figure \ref{CN_EXTN_FIT} uses an Orthogonal Distance Regression method, from the \texttt{scipy} package ODR, to ensure errors in [C/N] and log(Age) were both equally considered in our fit. Additionally, we did an investigation with a linear fit, however this fit resulted in too many stars being determined to be older than the age of the universe. The results of the linear fit can be found in the Appendix.

\section{Discussion}

To verify the reliability and consistency of our [C/N]-age relation, we need to compare to other independent well-regarded age determination methods, notably asteroseismology.

\subsection{Comparison to Asteroseismic Ages}
A natural comparison can be made between asteroseismic-based ages  from catalogs of evolved stars and our results. For this work, we compared to the APOKASC3 survey (Pinsonneault et al., {\it in press}) 
and the recently released APO-K2 \citep{APOK2} catalog. The APOKASC3 survey uses a combination of APOGEE and Kelper data to obtain stellar ages, while APOK-K2 uses a combination of APOGEE and K2 data \citep{APOK2_ages}. \citet{APOK2_ages} also recalculated ages for the APOKASC2 stars with their age determined method. The following sections detail our comparison to the two surveys and our verification of our [C/N]-ages.\\

\subsubsection{APOKASC3}

There are 8671 stars that meet our sample requirements and are also in APOKASC3. There are four open clusters and no globular clusters that are common between our sample and APOKASC3, but only three open clusters were found to have reliable cluster membership based on OCCAM probabilities; hence, only these three clusters were compared on a star-by-star basis.

\begin{figure}[h!]
\epsscale{1.1}
\plotone{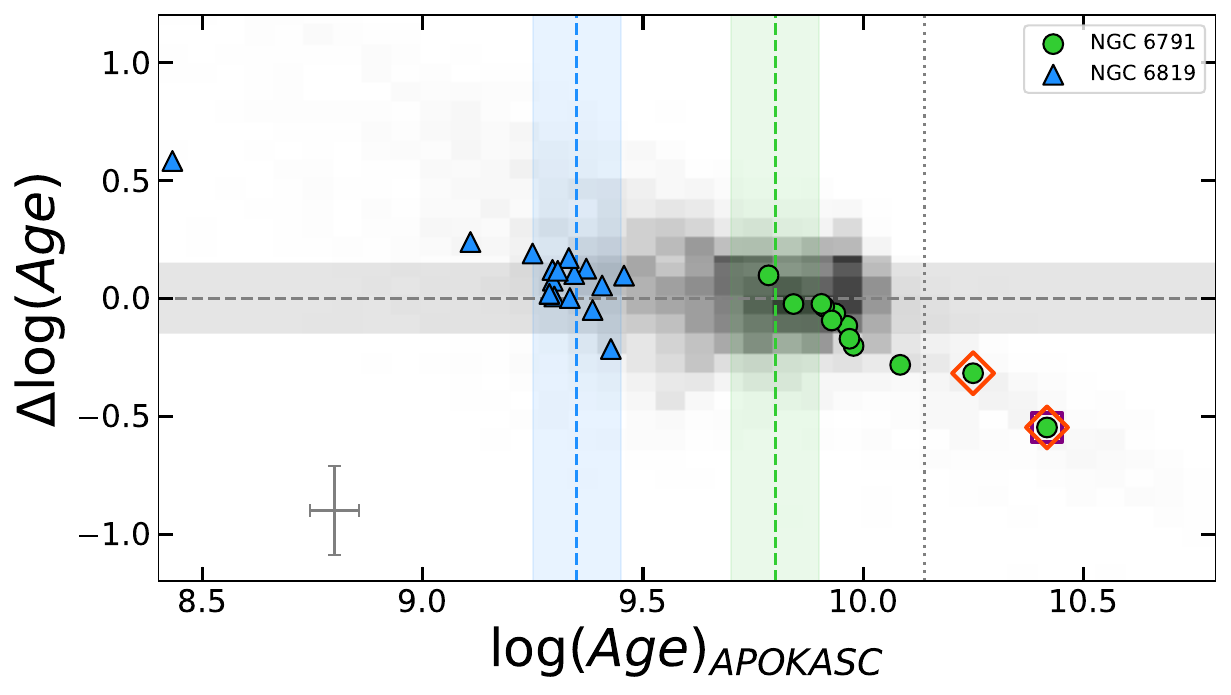}
\plotone{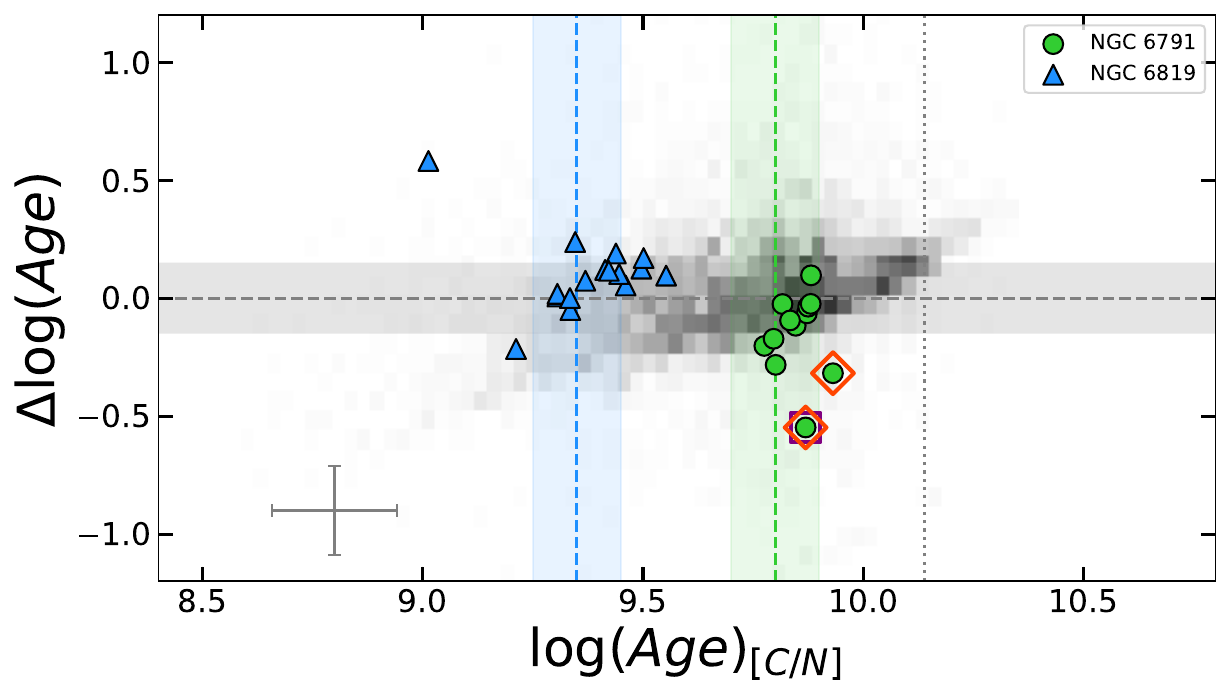}
\centering
\caption{\setlength{\baselineskip}{15.0pt} $\Delta \log(Age)$ as a function of APOKASC3 log(Age) ({\it top}) and [C/N]-based log(age) ({\it bottom}). The grey-gradient shaded regions represent bins of field stars; darker bins imply more stars. The green circles, blue triangles, and red diamonds represent cluster member stars common to both samples NGC 6791, NGC 6819, and NGC 6811, respectively. The vertical dashed lines are the \citet{cg20} determined ages for each cluster and the respective shaded regions shows the error in cluster age. The horizontal grey-dashed line is where $\Delta \log(Age) = 0$ with the surrounding grey region representing a delta of $-0.1$ and $+0.1$. The vertical grey-dotted line shows the currently accepted age of the universe  \citep[$13.8 \pm 0.05$ Gyr;][]{age1,Desi1a,Desi1b}. Median representative error bars are shown in each panel. Clusters member older than the universe are highlighted with an orange diamond were found to be rapidly rotating \citep{Patton2024_raprot} and those highlighted with a purple square were found to be in a binary system \citep{APW_bin}.}
\label{CN_APOKASC3_compare}
\end{figure}

For further verification, we apply our calibrations to open clusters that are also in the APOKASC3 sample: NGC 6791, NGC 6819, and NGC 6811, shown in Figure \ref{CN_APOKASC3_compare}. We find that our calculated ages are consistent with the ages determined in \citet{cg20} for all three clusters. {The spread in the [C/N]-calibration ages is due to the uncertainty of [C/N] in the individual cluster members, but such a spread is expected because our calibration is based on the average [C/N] abundance of the cluster. }

The outlier stars from NGC 6791 that are older than the universe, shown with green circles that are between the vertical grey dotted line and $\log(Age)=10.50$ highlighted in the top panel of Figure \ref{CN_APOKASC3_compare}, both are found to be in binary systems by \citet{APW_bin} while only one is found to be rapidly rotating by \citet{Patton2024_raprot}. The one found to be rapidly rotating is the green circle around $\log(Age)=10.25$ and just to the right of the vertical grey dotted line that represents the accepted age of the universe. Often such a star is associated with the wrong mass, chemistry, or age. This could be for physical reasons, e.g. 
that mass transfer or a tidal interaction has occurred \citep{Bufanda2023}, so the determined mass may not indicate age and consequently changing the surface abundance of carbon and nitrogen. 
However, it could also indicate offsets in the spectroscopic results. The APOGEE DR17 pipeline assumes giants are not significantly rotating. For rapidly rotating stars, this means the broadening effects that are not accounted for will change the observed temperature \citep{Dixon2020}, surface gravity \citep{Bufanda2023}, and much likely the carbon and nitrogen abundance. 

Stars that are calculated to be older than the universe only make up 3\% of our sample. About half of these stars we find to be rapidly rotating \citep{Patton2024_raprot} 
and/or in a binary system \citep{APW_bin} 
which could change the observed surface abundance of carbon and nitrogen \citep{Bufanda2023}.
Comparison with APOKASC3 results suggests that [C/N] based ages can be trusted to 10\% for 99.77\% of giants, as shown in Figure \ref{CN_APOKASC3_compare}.

\subsubsection{APO-K2}

\begin{figure}[ht!]
\epsscale{1.1}
\plotone{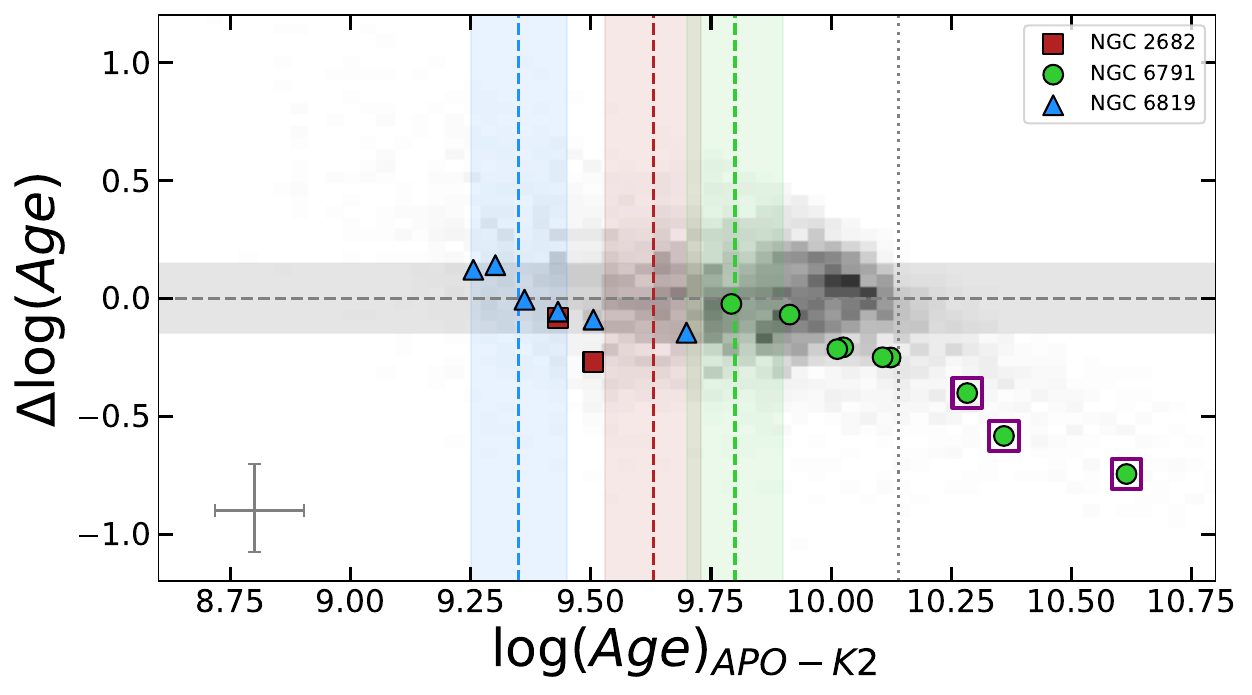}
\plotone{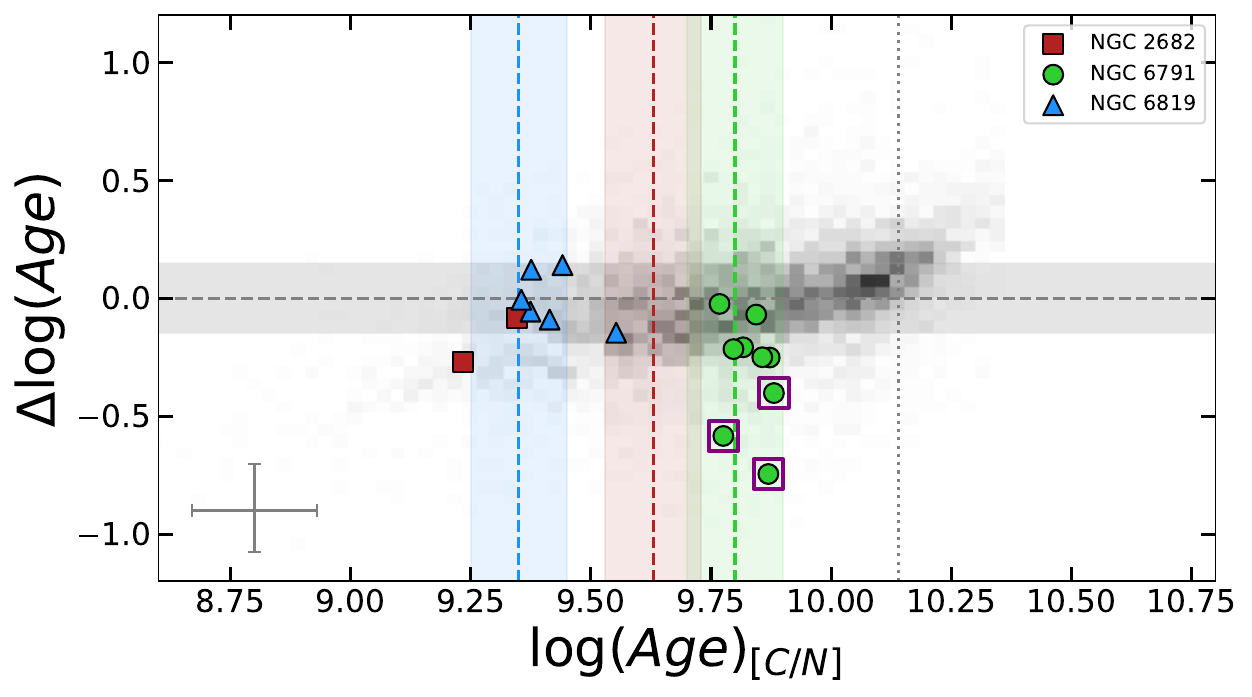}
\centering
\caption{\setlength{\baselineskip}{15.0pt} $\Delta \log(Age)$ as a function of APO-K2 log(Age) ({\it top}) and [C/N]-based log(age) ({\it bottom}). The grey-gradient shaded regions represent bins of field stars; darker bins imply more stars. The red squares, green circles, and blue triangles represent cluster member stars common to both samples NGC 2682 (M 67), NGC 6791, and NGC 6819, respectively. The vertical dashed lines are the \citet{cg20} determined ages for each cluster and the respective shaded regions shows the error in cluster age. The horizontal grey-dashed line is where $\Delta \log(Age) = 0$ with the surrounding grey region representing a delta of $-0.1$ and $+0.1$. The vertical grey-dotted line shows the accepted age of the universe. Median representative error bars are shown in each panel. Clusters member older than the universe are highlighted with a purple square were found to be in a binary system \citep{APW_bin}.}
\label{CN_APOK2_compare}
\end{figure}

Our sample shares 7120 stars in common with APO-K2, including 10 open clusters (but no globular clusters), but only 3 open cluster meet our seletion criteria. APO-K2 combines K2 asteroseismology, APOGEE spectroscopy, and Gaia astrometry and produced a catalog of 7672 evolved stars (either red giant branch or red clump) that have precise asteroseismic ages, masses, radii, abundance values and kinematic parameters. Ages for APO-K2 were derived in the second catalog release in \citet{APOK2_ages}. Of the three open clusters we compare ages on a star-by-star basis.

For further verification, we apply our calibrations to open clusters that are also in the APO-K2 sample, NGC 2682 (M 67) and the APO-K2 recalibrated APOKASC 2 sample, NGC 6791 and NGC 6819, as shown in Figure \ref{CN_APOK2_compare}. We find that our calculated ages are more consistent with the ages determined in \citet{cg20} for two of the three clusters than APO-K2. For both, APO-K2 and this work, the ages of stars within NGC 2682 (M 67) are underestimated relative to the \citet{cg20} age. While the APOKASC ages were consistent with the cluster ages and the [C/N] based ages for each cluster, we see slightly less agreement between the APO-K2 results and the cluster ages, particularly for stars in the recalibrated cluster NGC 6791.

From Figure \ref{CN_EXTN_FIT}, the cluster NGC 2682 is further off the linear fit than the other APO-K2 clusters used in our comparison and explains why we see a discrepancy between our derived ages and the \citet{cg20} ages used in our calibration. This discrepancy could be due to either additional effects in the APOGEE measurement of [C/N] (most NGC 2682 stars are significantly higher $S/N$ which many reveal weaker lines) or there may be a systematic error in the isochrone fitting age determination of NGC 2682 by \citet{cg20}, similar to as is seen in Figure 6 of \citet{Hunt2023} where blue stragglers affect the main sequence turn off location for older clusters from their work as well as \citet{cg20} and \citet{mwsc_catalog}.

In this comparison to APO-K2, we find that stars that are calculated to be older than the universe now comprise 9\% of our sample, and at least a third of those stars are clear radial velocity binary systems \citep{APW_bin} which could change the observed surface abundance of carbon and nitrogen \citep{Bufanda2023}. The stars determined to be ``older than the universe" as determined by the APO-K2 catalog, were checked for binarity in common proper motion (using Gaia DR3) and assessed for unresolved binaries using astrometric errors and markers (via priv. comm. regarding Schonhut-Stasik in prep.).
Similarly, the spread in the [C/N]-calibration ages is due to the uncertainty of [C/N] in the individual cluster members, but such a spread is expected because our calibration is based on the average [C/N] abundance of the cluster, with representative error shown.
Comparison with APO-K2 results suggests that [C/N] based ages are consistent with asterosiesmic ages to 10\% for 99.6\% of giants, as shown in Figure \ref{CN_APOK2_compare}.

\section{Conclusion}

This calibration, based on APOGEE DR17, provides a relation for RGB stars that experience the first dredge-up, but have not experienced extra-mixing (i.e., APOGEE DR17 stars with $\log g < 3.2$ but greater than the metallicity dependant extra-mixing values from Table 2 of \citet{Shetrone_2019}) as shown by Equation \ref{extn_cali_eqn}.
This relationship can be applied to the RGB stars within the metallicity range $-1.2\leq[Fe/H]\leq+0.3$ based on our sample.

Comparison with asteroseismic results of APOKASC and APO-K2 add additional evidencw that these [C/N] based ages are reliable. The APOKASC results suggest that [C/N] based ages can be trusted to 10\% for 99.77\% of giants, while APO-K2 results suggest that [C/N] based ages can be trusted to 10\% for 99.92\% of giants.

{The} majority of stars that are found to be older than the universe are either (1) rapidly rotating, (2) in a binary system, or (3) a combination of the two. As this relationship is being applied to a greater sample, keep in mind there are these potential exceptions to the relationship.

We can age-date over 45\% of the SDSS/APOGEE  DR17 sample, a total of over 300,000 stars.
This is exciting as we can gain a better insight into more regions of the Galaxy, as we are no longer restricted to the metal-rich disk, and paint a more coherent picture of our Galaxy's evolution.

\begin{acknowledgements}

TS, PMF, NM, JO, AW, and JD acknowledge support for this research from the National Science Foundation Astronomy and Astrophysics grants AST-1715662 and AST-2206541.  
JT and PMF acknowledge part of this work was performed at the Aspen Center for Physics, which is supported by National Science Foundation grant PHY-1607611.
KT and ETT gratefully acknowledges support for this research from the National Science Foundation’s Research Experience for Undergraduates program (PHY-1852267, PHY-2244258).
JSS would like to thank the Frist Center for Autism and Innovation in the School of Engineering at Vanderbilt University, who provide the Neurodiversity Inspired Science and Engineering Graduate Fellowship.
\end{acknowledgements}

Funding for the Sloan Digital Sky 
Survey IV has been provided by the 
Alfred P. Sloan Foundation, the U.S. 
Department of Energy Office of 
Science, and the Participating 
Institutions. 

SDSS-IV acknowledges support and 
resources from the Center for High 
Performance Computing  at the 
University of Utah. The SDSS 
website is www.sdss.org.

SDSS-IV is managed by the 
Astrophysical Research Consortium 
for the Participating Institutions 
of the SDSS Collaboration including 
the Brazilian Participation Group, 
the Carnegie Institution for Science, 
Carnegie Mellon University, Center for 
Astrophysics | Harvard \& 
Smithsonian, the Chilean Participation 
Group, the French Participation Group, 
Instituto de Astrof\'isica de 
Canarias, The Johns Hopkins 
University, Kavli Institute for the 
Physics and Mathematics of the 
Universe (IPMU) / University of 
Tokyo, the Korean Participation Group, 
Lawrence Berkeley National Laboratory, 
Leibniz Institut f\"ur Astrophysik 
Potsdam (AIP),  Max-Planck-Institut 
f\"ur Astronomie (MPIA Heidelberg), 
Max-Planck-Institut f\"ur 
Astrophysik (MPA Garching), 
Max-Planck-Institut f\"ur 
Extraterrestrische Physik (MPE), 
National Astronomical Observatories of 
China, New Mexico State University, 
New York University, University of 
Notre Dame, Observat\'ario 
Nacional / MCTI, The Ohio State 
University, Pennsylvania State 
University, Shanghai 
Astronomical Observatory, United 
Kingdom Participation Group, 
Universidad Nacional Aut\'onoma 
de M\'exico, University of Arizona, 
University of Colorado Boulder, 
University of Oxford, University of 
Portsmouth, University of Utah, 
University of Virginia, University 
of Washington, University of 
Wisconsin, Vanderbilt University, 
and Yale University.

This work has made use of data from the European Space Agency (ESA) mission {\it Gaia} (\url{https://www.cosmos.esa.int/gaia}), processed by the {\it Gaia} Data Processing and Analysis Consortium (DPAC, \url{https://www.cosmos.esa.int/web/gaia/dpac/consortium}). Funding for the DPAC has been provided by national institutions, in particular the institutions participating in the {\it Gaia} Multilateral Agreement.

This research made use of Astropy, a community-developed core Python package
    for Astronomy \citep{astropy:2013,astropy:2018}.

\facilities{Du Pont (APOGEE), Sloan (APOGEE), Spitzer, WISE, 2MASS, Gaia}
\software{\href{http://www.astropy.org/}{Astropy}~\citep{astropy:2013,astropy:2018}, \href{https://scipy.org}{SciPy}~\citep{SciPy}, \href{https://numpy.org}{NumPy}~\citep{Numpy}}

\bibliography{Spoo}{}
\bibliographystyle{aasjournal}

\appendix \label{appx_pipelines}
\section{[C/N]-Age Linear Relationship}

In this appendix, we present, for completeness, a linear fit to the cluster sample, which we chose not to use due producing numerous stars that would be deemed older than the age of the universe.

\subsection{Linear Fit Results}

After application of these additional membership and cluster age cuts the sample used for our [C/N]-age calibration (Figure ~\ref{CN_EXTN_FIT}) is comprised of the 49 clusters (530 stars) from \citet{occam_p7} plus four globular clusters: 47 TUC, M 71, M 4, and M 5 (31 stars). The full sample covers an age range of $8.62 \leq \log(Age[{\rm yr}]) \leq 10.13$ and a metallicity range of $-1.2 \leq {\rm [Fe/H]} \leq +0.3$. 
The individual globular cluster stars used in the calibration extension are shown in Table \ref{tab:GC_stars}. 

\subsubsection{The Extended DR17 [C/N] Abundance/Age Calibration}

In log-log space, assuming that the relationship between stellar age and [C/N] {is linear; our best fit is given by} 
{\small
\begin{equation}
{\log[Age({\rm yr})]_{\rm DR17}} = 10.23 \, (\pm 0.08) + 2.48\,(\pm 0.20) \, {\rm [C/N]}
\label{extn_cali_LINeqn}
\end{equation}
}
{and yields a Pearson coefficient of $R=0.86$.} 
Our Pearson coefficient is comparable to that found by \citet{casali_2019}{($R=0.85$)}, although we compute a slightly different offset and a weaker slope.
The linear fit given by Equation \ref{extn_cali_eqn} and shown in Figure \ref{CN_EXTN_FIT} uses an Orthogonal Distance Regression (ODR) method, from the \texttt{scipy} package ODR, to ensure errors in [C/N] and log(Age) were both equally considered in our fit.

\begin{figure*}[ht!]
\epsscale{1.15}
\plotone{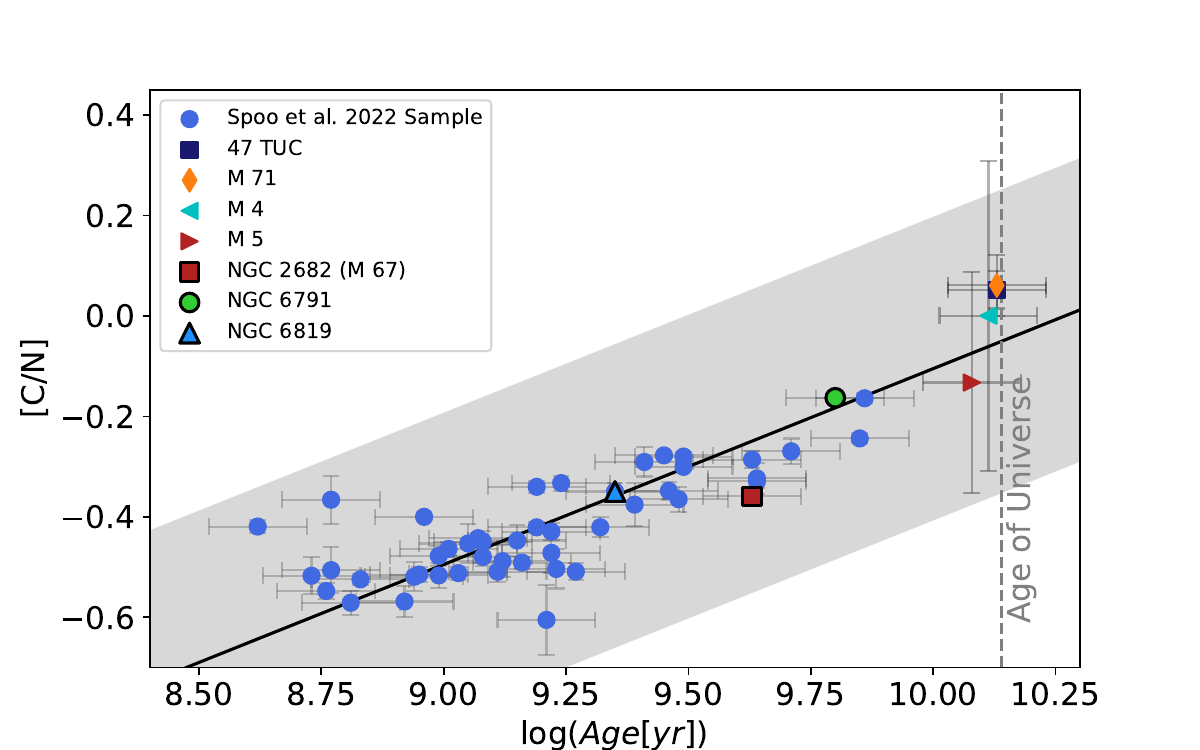}
\centering
\caption{\setlength{\baselineskip}{15.0pt} The [C/N] versus $\log(Age[{\rm yr}])$ distribution for the final sample, composed of \citet{occam_p7} open cluster sample and globular clusters: 47 TUC, M 71, M 4, and M 5. Clusters represented by blue circles are from the final calibration sample of \citet{occam_p7}. Globular clusters 47 TUC, M 71, M 4, M 5, are represented with an indigo square, an orange thin-diamond, a cyan left-triangle, and a red right-triangle, respectively. Our new linear fit is shown as a solid black line, and the fit's error envelope is shown with the grey region. The accepted age of the universe is shown as a vertical grey-dashed line. The open clusters used in our asteroseismic age comparison are NGC 2682 (M 67), NGC 6791, and NGC 6819 and are shown in black outlined red square, green circle, and blue triangle, respectively.}
\label{CN_EXTN_LINFIT}
\end{figure*}

\subsection{Comparison to Asteroseismic Ages}
A natural comparison can be done to asterosiesmic-based ages. For this work we compared to the APOKASC3 survey \citep[][Pinsonneault et al. {\it in press}]{apokasc}{}{}
and the recently released APO-K2 \citep{APOK2}. The APOKASC3 survey uses a combination of APOGEE and Kelper data to obtain stellar ages while APOK-K2 uses a combination of APOGEE and K2 data \citep{APOK2_ages}. \citet{APOK2_ages} also recalculated ages for the the APOKASC2 stars with their age determined method. The following sections details our comparison to the two surveys and our verification of our [C/N]-ages assuming the linear relation.

\subsubsection{APOKASC3}

\begin{figure}[ht!]
\epsscale{1.}
\plotone{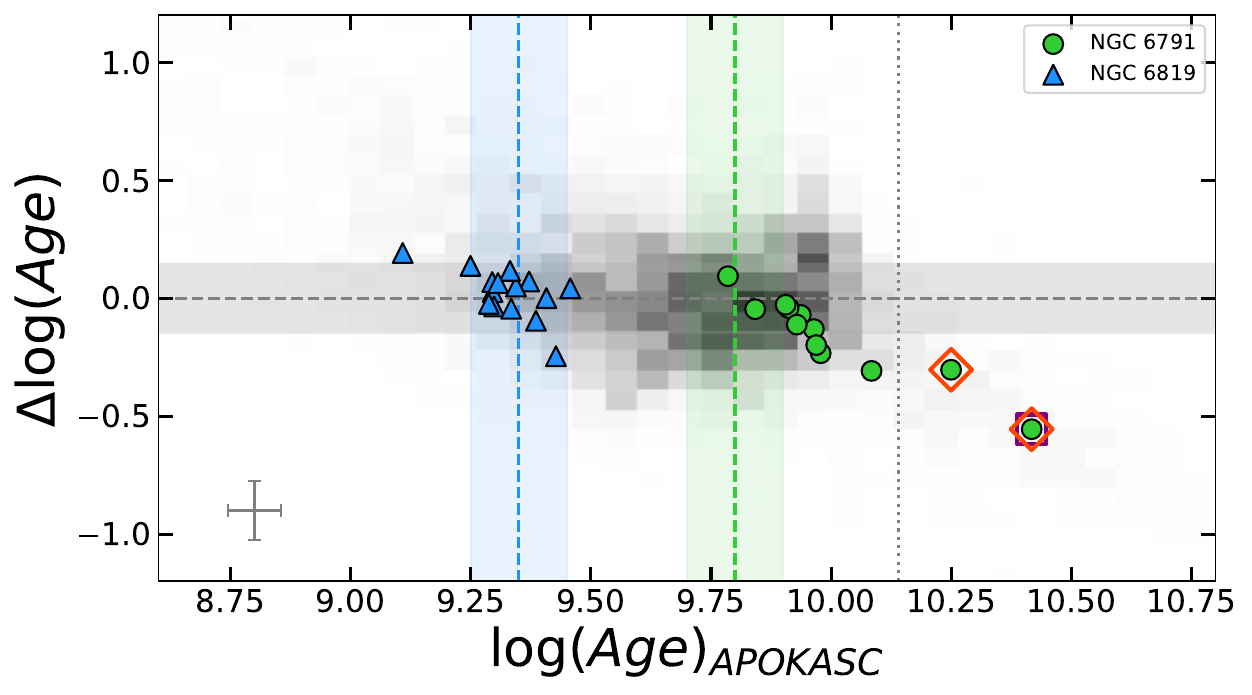}
\plotone{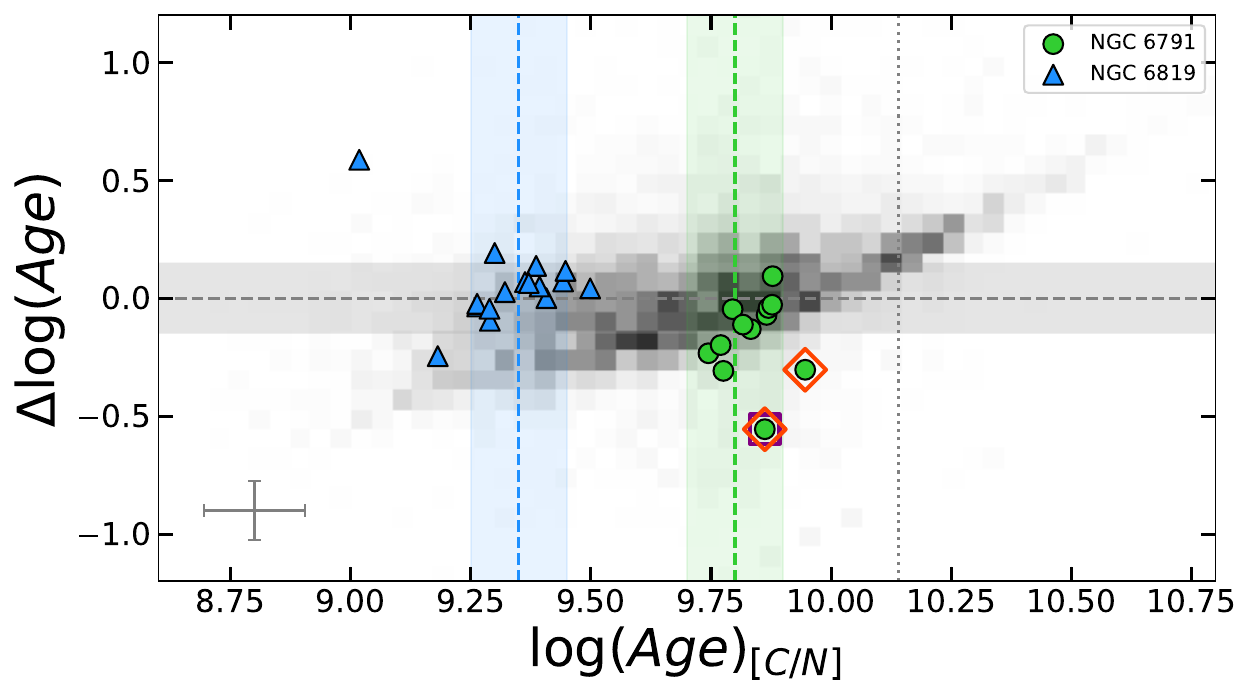}
\centering
\caption{\setlength{\baselineskip}{15.0pt} $\Delta \log(Age)$ as a function of APOKASC3 log(Age) ({\it top}) and [C/N]-based log(age) ({\it bottom}). The grey-gradient shaded regions represent bins of field stars; darker bins imply more stars. The green circles and blue triangles represent cluster member stars common to both samples NGC 6791 and NGC 6819, respectively. The vertical dashed lines are the \citet{cg20} determined ages for each cluster and the respective shaded regions shows the error in cluster age. The horizontal grey-dashed line is where $\Delta \log(Age) = 0$ with the surrounding grey region representing a delta of $-0.1$ and $+0.1$. The vertical grey-dotted line shows the accepted age of the universe. Median representative error bars are shown in each panel. Clusters member older than the universe are highlighted with an orange diamond were found to be rapidly rotating \citep{Patton2024_raprot} and those highlighted with a purple square were found to be in a binary system \citep{APW_bin}.}
\label{CN_APOKASC3_LINcompare}
\end{figure}

There are 6004 stars that meet our criteria that are also in APOKASC3. There are four open clusters and no globular clusters that are common between our sample and APOKASC3, but only three open clusters were found to have reliable cluster membership based on OCCAM probabilities, hence we compare only these three clusters were compared on a star-by-star basis.

For further verification, we apply our calibrations to open clusters that are also in the APOKASC sample: NGC 6791 and NGC 6819, shown in Figure \ref{CN_APOKASC3_compare}. We find that our calculated ages from the DR17 calibration are consistent to the ages determined in \citet{cg20} for all three clusters. The spread in the [C/N]-calibration ages within the cluster is due to the uncertainty of [C/N] in the individual cluster members, but such a spread is expected because our calibration is based on the average [C/N] abundance of the cluster.

The outlier stars from NGC 6791 that are older than the universe, shown with green circles that are between the vertical grey dotted line and $\log(Age)=10.50$ in the bottom panel of Figure \ref{CN_APOKASC3_LINcompare}, both are found to be in binary systems by \citet{APW_bin} while only one is found to be rapidly rotating by \citet{Patton2024_raprot}. The one found to be rapidly rotating is the green circle around $\log(Age)=10.25$ and just to the right of the vertical grey dotted line that represents the accepted age of the universe. Since the star is rapidly rotating, this usually indicates that mass transfer or tidal interactions has occurred, so the determined mass may not indicate age as well as potentially changing the surface abundance of carbon and nitrogen. Another thing to note for the rapidly rotating stars, the APOGEE DR17 pipeline assumes giants are not significantly rotating and the broadening effects that are not accounted for will change the observed carbon and nitrogen abundance.

Stars that are calculated to be older than the universe only make up 4\% of our sample, where about third of those stars we find them to be rapidly rotating \citep{Patton2024_raprot} and/or in a binary system \citep{APW_bin} which could change the observed relationship between age and the surface abundance of carbon and nitrogen \citep{Bufanda2023}.
Comparison with APOKASC results suggests that [C/N] based ages can be trusted to 10\% for 98.1\% of giants using the linear relation.

\subsubsection{APO-K2}

\begin{figure}[ht!]
\epsscale{1.}
\plotone{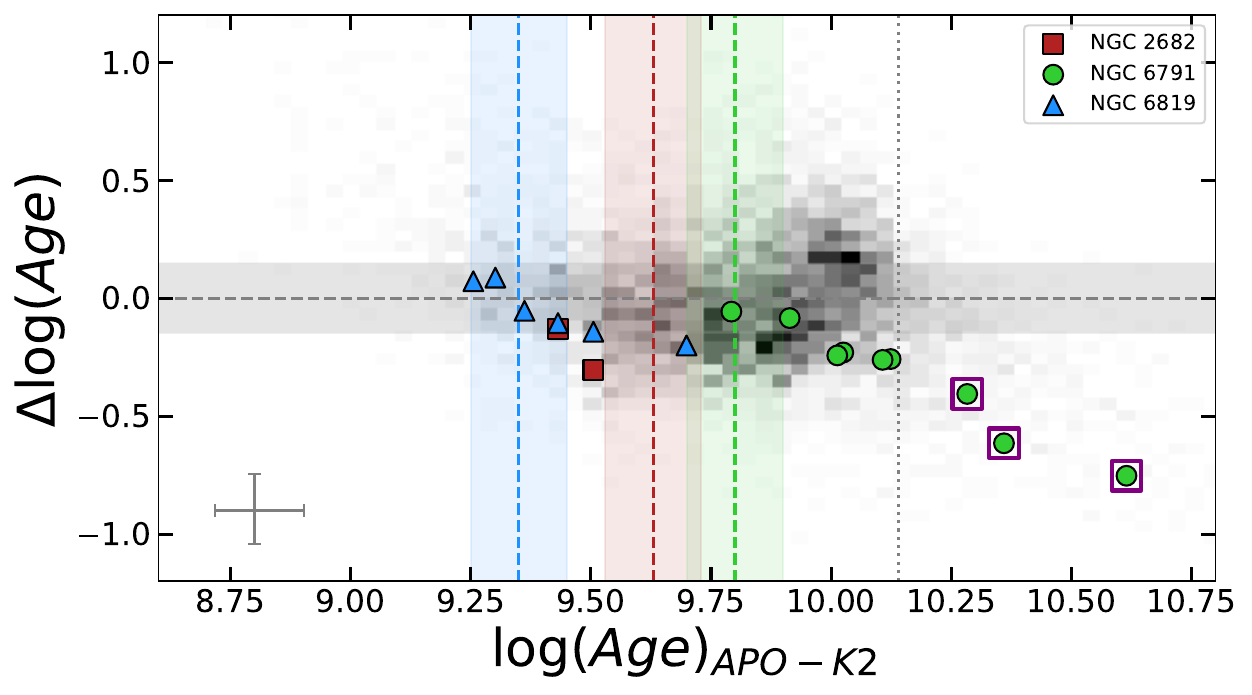}
\plotone{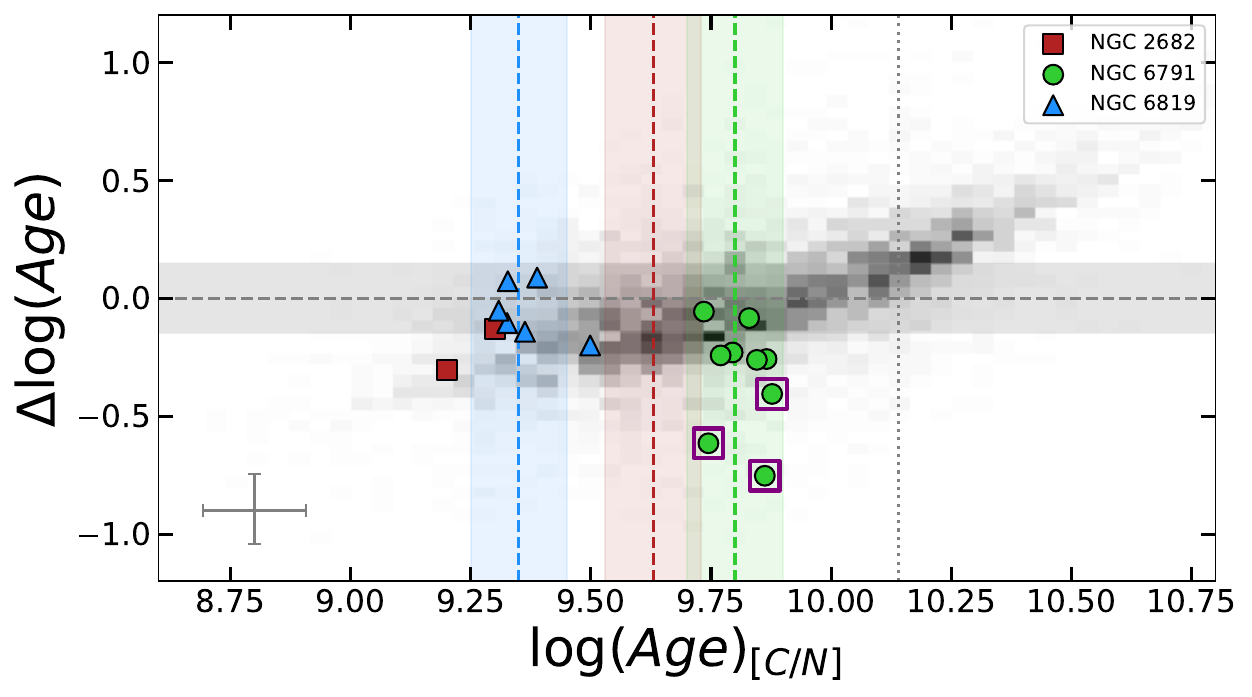}
\centering
\caption{\setlength{\baselineskip}{15.0pt} $\Delta \log(Age)$ as a function of APO-K2 log(Age) ({\it top}) and [C/N]-based log(age) ({\it bottom}). The grey-gradient shaded regions represent bins of field stars; darker bins imply more stars. The red squares, green circles, and blue triangles represent cluster member stars common to both samples NGC 2682 (M 67), NGC 6791, and NGC 6819, respectively. The vertical dashed lines are the \citet{cg20} determined ages for each cluster and the respective shaded regions shows the error in cluster age. The horizontal grey-dashed line is where $\Delta \log(Age) = 0$ with the surrounding grey region representing a delta of $-0.1$ and $+0.1$. The vertical grey-dotted line shows the accepted age of the universe. Median representative error bars are shown in each panel. Clusters member older than the universe are highlighted a purple square were found to be in a binary system \citep{APW_bin}.}
\label{CN_APOK2_LINcompare}
\end{figure}

There are 6881 stars in our sample that are also in APO-K2. There are 10 open clusters and no globular clusters that are common between our sample and APO-K2, but only three open clusters were found to have reliable cluster membership based on OCCAM probabilities, hence we compare only these three clusters were compared on a star-by-star basis.

For further verification, we apply our calibrations to open clusters that are also in the APO-K2 sample, NGC 2682 (M 67), and the APO-K2 recalibrated APOKASC 2 sample, NGC 6791 and NGC 6819, as shown in Figure \ref{CN_APOK2_LINcompare}. We find that our calculated ages from the DR17 calibration are more consistent with the ages determined in \citet{cg20} for three of the four clusters than APO-K2. For both, APO-K2 and this work, the ages of stars within NGC 2682 (M 67) are under estimated. 
From Figure \ref{CN_EXTN_LINFIT}, the cluster NGC 2682 is further off the linear fit than the other APO-K2 cluster used in our comparison and explains why we show a discrepancy between our derived ages and the \citet{cg20} ages used in our calibration. This discrepancy could be due to either additional effects in the APOGEE measurement of [C/N] (most NGC 2682 stars are significantly higher $S/N$ which many reveal weaker lines) or there may be a systematic in the isochrone fitting age determination of NGC 2682 by \citet{cg20}, similar to as is seen in Figure 6. of \citet{Hunt2023} where blue stragglers affect the main sequence turn off location for older clusters from their work as well as \citet{cg20} and \citet{mwsc_catalog}.
Stars that are calculated to be older than the universe only make up 20\% of our sample, where a third of those stars we find to be in a binary system \citep{APW_bin} which could change the observed surface abundance of carbon and nitrogen \citep{Bufanda2023}. Stars older than the universe that were present in the APO-K2 catalog, were checked for binarity in common proper motion (using Gaia DR3) and assess for unresolved binaries using astrometric errors and markers (via priv. comm. regarding Schonhut-Stasik in prep.).
Similarly, the spread in the [C/N]-calibration ages is due to the uncertainty of [C/N] in the individual cluster members, but such a spread is expected because our calibration is based on the average [C/N] abundance of the cluster.
Comparison with APO-K2 results suggests that [C/N] based ages can be trusted to 10\% for 99.5\% of giants, as shown in Figure \ref{CN_APOK2_LINcompare}, assuming the linear relationship.

\end{document}